\documentclass[aps,prl,twocolumn,floatfix,superscriptaddress,footinbib, sort&compress, numbers, merge, reprint]{revtex4-1}
\usepackage{graphicx}
\usepackage{amsmath}
\usepackage{amssymb}
\usepackage{booktabs}
\usepackage{color}
\usepackage{natbib}
\usepackage{hyperref}

\usepackage{chngcntr}

\newcommand{\abs}[1]{\lvert #1\rvert}

\usepackage[utf8]{inputenc}
\usepackage[tight,sf]{subfigure}%

\usepackage{dcolumn}
\usepackage{array}
\usepackage{tabularx}

\usepackage{floatrow}

\usepackage{eucal}
\usepackage[dvips]{epsfig}
\usepackage{amsfonts}

\usepackage{amsthm}

\newcommand{\be}{\begin{eqnarray}}
\newcommand{\ee}{\end{eqnarray}}
\newcommand{\bse}{\begin{subequations}}
\newcommand{\ese}{\end{subequations}}

\newcommand{\bnum}{\begin{enumerate}}
\newcommand{\enum}{\end{enumerate}}

\newcommand{\bit}{\begin{itemize}}
\newcommand{\eit}{\end{itemize}}

\newcommand{\bc}{\begin{cases}}
\newcommand{\ec}{\end{cases}}

\newcommand{\bpm}{\begin{pmatrix}}
\newcommand{\epm}{\end{pmatrix}}

\newcommand{\bvm}{\begin{vmatrix}}
\newcommand{\evm}{\end{vmatrix}}

\newcommand{\bs}{\boldsymbol}

\newcommand{\mrm}{\mathrm}

\newcommand{\Gl}{\Lambda}

\newcommand{\p}{\partial}
\newcommand{\f}{\frac}

\newcommand{\lan}{\langle}
\newcommand{\ran}{\rangle}

\newcommand{\tn}{\textnormal}

\begin{document}

\title{Anomalous chained turbulence in actively driven flows on spheres}

\author{Oscar Mickelin}
\affiliation{Department of Mathematics, Massachusetts Institute of Technology, Cambridge,~MA~02139-4307, USA}

\author{Jonasz S\l{}omka}
\affiliation{Department of Mathematics, Massachusetts Institute of Technology, Cambridge,~MA~02139-4307, USA}

\author{Keaton J. Burns}
\affiliation{Department of Physics, Massachusetts Institute of Technology, Cambridge,~MA~02139-4307, USA}

\author{\\Daniel~Lecoanet}
\affiliation{Princeton Center for Theoretical Science, Princeton University, Princeton, NJ~08544, USA}

\author{Geoffrey M. Vasil}
\affiliation{School of Mathematics and Statistics, University of Sydney,~NSW~2006, Australia}
	
\author{Luiz M. Faria}
\affiliation{Department of Mathematics, Massachusetts Institute of Technology, Cambridge,~MA~02139-4307, USA}

\author{J\"orn Dunkel} 
\affiliation{Department of Mathematics, Massachusetts Institute of Technology, Cambridge,~MA~02139-4307, USA}
\date{\today}
   
\begin{abstract}
Recent experiments demonstrate the importance of substrate curvature for actively forced fluid  dynamics. Yet, the covariant formulation and analysis of continuum models for non-equilibrium flows on curved surfaces still poses theoretical challenges. Here, we introduce and study a generalized covariant Navier-Stokes model for fluid flows driven by active stresses in non-planar geometries. The analytical tractability of the theory is demonstrated through exact stationary solutions for the case of a spherical bubble geometry. Direct numerical simulations reveal a curvature-induced transition from a burst phase to an anomalous turbulent phase that differs distinctly from externally forced classical  2D Kolmogorov turbulence.  This new type of active turbulence is characterized by the self-assembly of finite-size vortices into linked chains of anti-ferromagnetic order, which percolate through the entire fluid domain, forming an active dynamic network.  The coherent motion of the vortex chain network provides an efficient mechanism for upward energy transfer from smaller to larger scales, presenting an alternative to the conventional energy cascade in classical 2D turbulence. 
\end{abstract}
 

\maketitle


Substrate geometry profoundly affects dynamics and energy transport in complex fluids flowing far from equilibrium~\cite{2005Cranmer,keber2014topology,Zhang:2016aa}.  Examples range from magnetohydrodynamic turbulence on stellar surfaces~\cite{2005Cranmer} to the rich microscale dynamics of topological defects in active nematic vesicles \cite{keber2014topology,Zhang:2016aa}. Studying the interplay between spatial curvature and actively driven fluid flows is also essential for understanding microbial locomotion~\cite{sipos2015hydrodynamic}, biofilm formation~\cite{2015Chang_NJP} and bioremediation~\cite{Das2007} in soils~\cite{Bold:1949aa}, tissues~\cite{Costerton:1999aa} and water~\cite{1992Rosenberg,2006Rosenberg,2013Rosenberg}. 
Over the past two decades, important breakthroughs have been made in characterizing active-stress driven  matter flows in planar Euclidean geometries both theoretically~\cite{vicsek1995novel,toner1995long,baskaran2009statistical,koch2011collective,ramaswamy2010mechanics,marchetti2013hydrodynamics,kruse2004asters,wolgemuth2008collective,saintillan2008instabilities,peshkov2012nonlinear,brotto2013hydrodynamics,thampi2013velocity} and experimentally~\cite{mendelson1999organized,steager2008dynamics,zhang2010upper,sokolov2007concentration,dombrowski2004self,wu2006collective,dunkel2013fluid}. More recently, theoretical work has begun to focus on incorporating curvature effects into active  matter models~\cite{sknepnek2015active,fily2016active,janssen2017aging,shankar2017topological,salbreux2017mechanics,henkes2017dynamical,duan2017curvature,fily2014dynamics,fily2015dynamics,malgaretti2017model}. Despite some promising progress, the hydrodynamic description of pattern-forming non-equilibrium  liquids in non-Euclidean spaces continues to pose conceptual challenges,  attributable to the difficulty of formulating exactly solvable continuum models and devising efficient spectral methods in curved geometries.
\par
Aiming to help improve upon these two issues, we introduce and  investigate here the covariant extension of a generalized Navier-Stokes (GNS) model~\cite{1993BeNi_PhysD,1996Tribelsky_PRL,slomka2017geometry,slomka2017spontaneous} describing incompressible active fluid flow on an arbitrarily curved surface.  Focusing on a spherical \lq bubble\rq~geometry, we derive  exact stationary solutions and numerically explore the effects of curvature on the steady-state flow dynamics, using the open-source spectral code Dedalus~\cite{dedalus2017}. The numerically obtained phase diagrams, energy spectra and flux curves predict  an anomalous turbulent phase when the spectral bandwidth of the active stresses becomes sufficiently narrow. This novel type of 2D turbulence supports an unexpected upward energy transfer mechanism, mediated by the large-scale collective dynamics of self-organized vortex chains, akin to actively moving anti-ferromagnetic spin chains.    At high curvature, the anomalous turbulence transforms into a quasi-stationary burst phase, whereas for broadband spectral forcing the flow dynamics transitions to classical 2D Kolmogorov turbulence, accumulating energy in a few large-scale vortices. We next motivate and define the covariant GNS model for an arbitrary 2D surface; analytical and numerical results for the sphere case will be discussed subsequently. 
\par
 Recent experiments have investigated the collective dynamics of swimming bacteria~\cite{sokolov2007concentration} and algae~\cite{2010Guasto} in thin quasi-2D soap films held by a coplanar wire frame. Generalizing to non-Euclidean geometries~\cite{keber2014topology,Zhang:2016aa}, which can be realized with soap bubbles or curved wire frames~\cite{Goldstein10062014}, we consider here a  free-standing non-planar 2D film in which the fluid flow is driven by active stresses, as in suspensions of swimming bacteria~\cite{2009LaugaPowers,2011DrescherEtAl} or ATP-driven microtubule networks~\cite{ramaswamy2010mechanics,2012Sanchez_Nature}. Assuming  incompressibility, the fluid velocity field components $v^a(t,\bs x)$,  obey the Cauchy momentum equation on a curved manifold~\cite{scriven1960dynamics,aris1989vectors},
\bse
\label{e:eom}
\be
\nabla_a v^a&=&0, \label{e:eom1}\\
\p_tv^a+v^b \nabla_b v^a&=&\nabla^a \sigma +\nabla_b T^{ab},\label{e:eom2}
\ee
where $\nabla_b v^a$ denotes the covariant derivative of $v^a$, $a,b=1,2$ and $\sigma$ is the (surface) tension. The stress tensor $T^{ab}$ includes passive and active contributions from the solvent fluid viscosity and the stresses exerted by the microswimmers on the fluid.  Below, we study the covariant version of the linear active-stress model~\cite{1993BeNi_PhysD,1996Tribelsky_PRL,slomka2017geometry,slomka2017spontaneous} 
\be
\label{eq:stress_tensor}
T^{ab}&=&f(\nabla^2)(\nabla^au^b+\nabla^bu^a), \\
f(\nabla^2)&=&\Gamma_0-\Gamma_2 \nabla^2+\Gamma_4\nabla^2\nabla^2,
\notag
\ee
\ese
where $\nabla^2=\nabla^a\nabla_a$ is the tensor Laplacian. In qualitative agreement with experimental observations for active suspensions~\cite{dombrowski2004self,sokolov2007concentration,dunkel2013fluid,2012Sanchez_Nature}, the polynomial ansatz for $f$ in Eq.~(\ref{eq:stress_tensor}) generates vortices of characteristic size $\Lambda$ and growth time $\tau$, provided that $\Gamma_2<0$, which introduces a bandwidth $\kappa$ of linearly unstable modes~\cite{slomka2017geometry}. General mathematical stability considerations demand $\Gamma_0,\Gamma_4>0$. The phenomenological model~\eqref{e:eom}  is minimal in the sense that it assumes the active stresses create to leading order a linear instability, while neglecting energy transfer within the active component. As verified in Ref.~\cite{slomka2017spontaneous}, the linear active-stress model~\eqref{eq:stress_tensor} suffices to quantitatively reproduce  the experimentally measured velocity distributions and flow correlations in 3D bacterial~\cite{dunkel2013fluid} and ATP-driven microtubule~\cite{2012Sanchez_Nature} suspensions. More generally,  closely related GNS models have also been studied in the context of soft-mode turbulence and seismic waves~\cite{1993BeNi_PhysD,1996Tribelsky_PRL} and the resulting non-equilibrium flow patterns share significant phenomenological similarities with  magneto-hydrodynamic (MHD) flows driven by electromagnetic stresses~\footnote{G. M. Vasil and M. G. P. Cassell, in preparation.}, suggesting that our results may have implications beyond microbial suspensions.

\par
\begin{figure}[t!]
\includegraphics[width=1\textwidth]{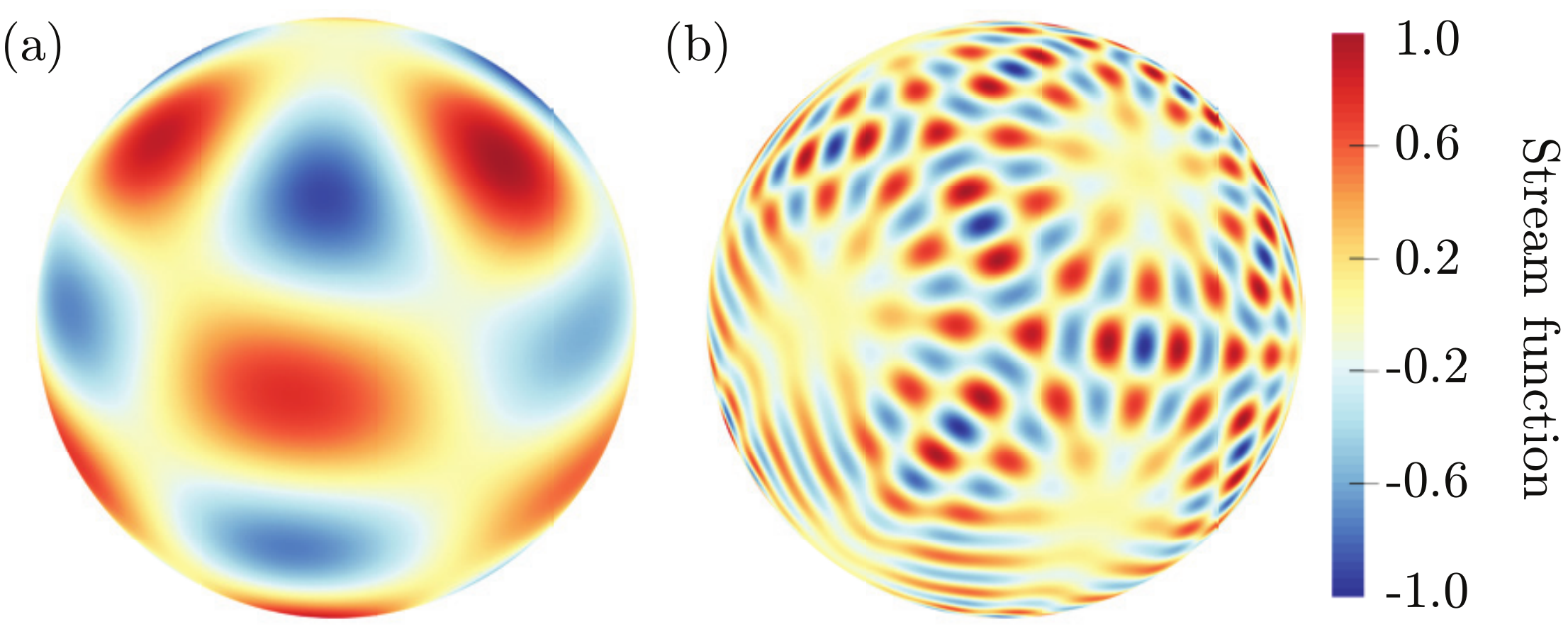}
\caption{Stationary solutions of Eqs.~\eqref{e:omegapsi} are superpositions of the form~\eqref{e:psi} with $f(-\ell(\ell+1)+4)=0$.
(a)~An exact stationary solution with $\ell=6$  which is also approximately realized as a transient state in the time-dependent burst solution of Fig.~\ref{fig:fig2} (Movie~1).
(b)~Complex symmetric solutions can be constructed by choosing the expansion coefficients $\psi_{m\ell}$ accordingly~\cite{prandl1996recursive}. In both panels, the stream functions are normalized by their maxima; see SM~\cite{SM} for coefficients $\psi_{m\ell}$. }
\label{fig:fig1}
\end{figure}
\begin{figure*}[t!!!!!]
\includegraphics[width=0.95\textwidth]{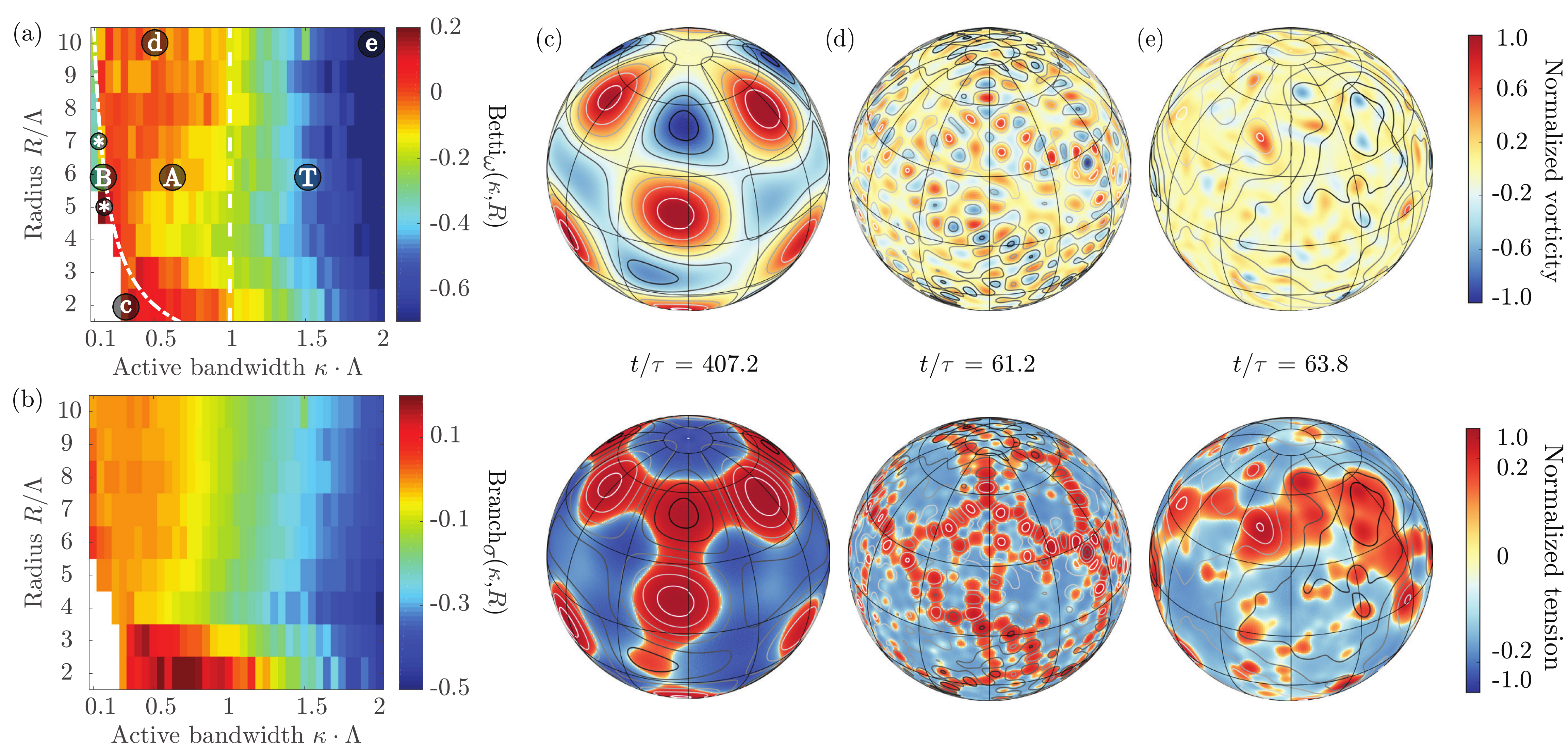}
\caption{Phase diagrams (a,b) and representative still images (c-e) from simulations showing quasi-stationary burst dynamics (B-phase), anomalous vortex-network turbulence~(A-phase), and classical 2D turbulence~(T-phase).
(a,b) The A- and T-phase are approximately separated by the condition $\kappa \Lambda=1$ (vertical dashed line) and differ by the average number of vortices (a), the branch geometry of the tension field (b), and the energy spectra (Fig.~\ref{fig:fig3}).  The B-phase arises for narrowband energy injection $\kappa R \lesssim 1$ when only a single $\ell$-mode is active (region right below the dashed-dotted line); decreasing $\kappa$ further gives a passive fluid (white region).
(c-e)~Top: Instantaneous vorticity fields normalized by their maxima. 
Bottom: Surface tension fields normalized by the maximum deviation from the mean.
(c) Quasi-stationary pre-burst state from Movie~1 resembling the exact solution in Fig.~\ref{fig:fig1}(a); see Movies~2 and 3 for additional examples labeled by $\ast$ in panel~(a).
(d) For subcritical curvature and intermediate energy injection bandwidths, $R^{-1}<\kappa<\Lambda^{-1}$, the flows develop a percolating vortex-chain network structure (Movie~4), with accumulation of tension and vorticity along the edges. 
(e) For broadband energy injection $\kappa\Lambda>1$, smaller eddies merge to create larger vortices, as typical of classical 2D turbulence (Movie~5).  Parameters: 
(a)~$\alpha_{\omega}=0.5$; 
(c)~$R/\Lambda = 2$, $\tau = 4.9 \text{ s}$, $\kappa \Lambda = 0.29$;
(d)~$R/\Lambda = 10$,   $\tau = 14.9 \text{ s}$, $\kappa \Lambda = 0.5$; 
(e)~$R/\Lambda = 10$,   $\tau = 11.7 \text{ s}$, $\kappa \Lambda = 2.0$. 
Panels (a, b) show steady-state time averages over~$[50\tau, 100\tau]$. Solid curves in (c-e)  indicate stream lines of the velocity fields. }
\label{fig:fig2}
\end{figure*}

Exact stationary solutions of Eqs.~(\ref{e:eom}) for a sphere of radius $R$ can be constructed from the equivalent vorticity-stream function formulation~(SM~\cite{SM})
\bse
\label{e:omegapsi}
\be
\Delta \psi&=&-\omega, \\
\p_t \omega+\{\omega,\psi\}&=&
f(\Delta+4K)(\Delta +2K)\omega,
\ee
\ese
where $\psi$ and $\omega$ are the stream function and vorticity.  The advection term in spherical coordinates $(\theta,\phi)$ reads $\{\omega,\psi\}=(\p_\theta\omega\p_\phi\psi -\p_\phi\omega \p_\theta\psi )/(R^2\sin\theta)$, the Gaussian curvature is $K=1/R^{2}$, and $\Delta$ denotes the standard spherical Laplacian. Since the spherical harmonics $Y^m_\ell$ diagonalize the Laplacian, $\Delta Y^m_\ell=-R^{-2}\ell(\ell+1)Y^m_\ell$, for integers $\ell,\,m$ such that $\ell\geq 0$ and $|m|\leq \ell$, an arbitrary superposition
\be\label{e:psi}
\psi=\sum_{|m|\leq \ell}\psi_{m\ell}Y^m_\ell
\ee
solves the system~(\ref{e:omegapsi}) exactly, provided that the eigenvalue $\ell$ is an integer root of
$f(-\ell(\ell+1)+4)=0$ (SM~\cite{SM}). As usual, the velocity field is tangent to the level sets of the stream function. Two particular exact solutions are shown in Fig.~\ref{fig:fig1}. The first example, Fig.~\ref{fig:fig1}(a), is reminiscent of the square lattice solutions found earlier in the flat 2D case~\cite{slomka2017geometry}. The second example in Fig.~\ref{fig:fig1}(b) illustrates a flow field with five-fold symmetry, obtained by applying the superposition procedure of Ref.~\cite{prandl1996recursive}. Although these exact solutions are not stable, they provide some useful  intuition about the instantaneous flow patterns expected in dynamical simulations (Fig.~\ref{fig:fig2}), similar to exact coherent structures~\cite{2001Waleffe_JFM} in conventional turbulence~\cite{2009Waleffe}.

To find and analyze time-dependent solutions of Eqs.~\eqref{e:eom}, we performed numerical simulations  using Dedalus \cite{dedalus2017}, an open-source framework for solving differential equations with spectral methods. The equations~\eqref{e:eom} were solved directly as a coupled partial differential-algebraic system for the scalar tension $\sigma$ and vector velocity $v^a$. To spatially discretize the system, we used spin-weighted spherical harmonics, which are a parameterized family of basis functions that correctly capture the analytical behavior of spin-weighted functions on the sphere.
The components of any order tensor on the sphere, when written in the spinor basis $\mathbf{e}_\theta \pm i \mathbf{e}_\phi$, take on definite spin weight $s$ and are therefore ideally represented using the corresponding spin-weighted harmonics~(SM~\cite{SM}).
For scalar functions, $s = 0$ and the spin-weighted spherical harmonics reduce to the normal scalar spherical harmonics.
The scalar tension and vorticity fields are represented in $s=0$ spherical harmonics, and the vector velocity field is represented in $s=\pm 1$ harmonics, up to a predetermined degree $\ell_{\text{max}}$.

\par
Under this spectral expansion, the system~\eqref{e:eom} is reduced to a set of coupled ordinary differential-algebraic equations for the time evolution of the expansion coefficients. We solve these equations using mixed implicit-explicit timestepping, in which the linear terms of the evolution equations are integrated implicitly, the linear constraints are enforced implicitly, and the nonlinear terms are integrated explicitly.
This allows us to simultaneously evolve the velocity field while enforcing the incompressibility constraint, and with a timestep that is limited by the advective Courant-Friedrichs-Lewy time condition rather than the diffusive time at any scale. Since the equations are linearly decoupled for different values of $m$, the scheme can be easily parallelized. This is done automatically via MPI in Dedalus, allowing the simulations to be run on up to $\ell_{\text{max}}$ cores simultaneously.

\par
The parameters $(\Gamma_0, \Gamma_2, \Gamma_4)$ in Eqs.~\eqref{e:eom} define a characteristic time scale~$\tau$, a characteristic vortex diameter~$\Lambda$, and a characteristic spectral bandwidth $\kappa$, which can be directly inferred from experimental data~\cite{slomka2017spontaneous}; explicit expressions are derived in the SM~\cite{SM}. Given a sphere of radius $R$, fixing $(\tau,\Lambda,\kappa)$ uniquely determines the parameters $(\Gamma_0, \Gamma_2, \Gamma_4)$. To explore the interplay between curvature and activity, we run  $351$ simulations, using $R/\Lambda \in [2, 10]$ and $\kappa\cdot \Lambda \in [0.1, 2.0]$.  Typical vortex diameters for bacterial and microtubule suspensions are~$\Gl\sim 50-100\;\mu$m with $\tau$ of the order of seconds \cite{dombrowski2004self,sokolov2007concentration,2012Sanchez_Nature,dunkel2013fluid}. Time steps were in the range $[5\cdot 10^{-4}\tau , 5\cdot 10^{-3}\tau]$  with a total simulation time $100\tau$, allowing the system to fully develop its dynamics  after an initial relaxation phase during which active stresses inject energy until the viscous dissipation and activity balance on average. In the remainder,  it will be convenient to regard $\Gl$ as reference length and compare the flow topologies across the $(\kappa, R)$ parameter plane.

Our simulations reveal three qualitatively distinct flow regimes (Fig.~\ref{fig:fig2}):  a quasi-stationary burst phase for~\mbox{$\kappa R \lesssim1$} [domain B in Fig.~\ref{fig:fig2}(a); Movies~1-3], an anomalous turbulence for $R^{-1}<\kappa<\Lambda^{-1}$  [domain A in Fig.~\ref{fig:fig2}(a); Movie~4], and normal 2D turbulence  for~$\kappa \Lambda> 1$ [domain T in Fig.~\ref{fig:fig2}(a); Movie~5]. Representative vorticity and tension fields from the corresponding steady-state dynamics are shown in Fig.~\ref{fig:fig2}(c-e).
\par
In the  B-phase, the  energy injection bandwidth $\kappa$ is close to the wavenumber spacing set by the sphere curvature $R^{-1}$, leaving only a single active wavenumber $\ell$.  Decreasing  $\kappa$ further  completely suppresses active modes resulting in globally damped fluid motion [white domain in Fig.~\ref{fig:fig2}(a)]. The B-phase is characterized by the formation of intermittent quasi-stationary flow patterns that lie in the vicinity of the  exact stationary solutions~(\ref{e:psi}), cf. Figs.~\ref{fig:fig1}(a) and~\ref{fig:fig2}(c).  Once formed, the amplitude of these flow patterns grows exponentially (Fig.~S3) until nonlinear advection becomes dominant and eventually causes energy to be released through a  rapid burst. Afterwards, the dynamics becomes quasi-linear again with the flow settling into a new quasi-stationary pattern. These burst cycles are continuously repeated~(Movies 1-3).
\par
The two turbulent phases  A and T in Fig.~\ref{fig:fig2}(a) can be distinguished through topological, geometric and spectral measures. We demonstrate this by determining  the topology of the vorticity fields,  the geometry of the high-tension domains and the energy spectra for each simulation after flows had reached the chaotic steady-state.  
\par 
To study the vortex topology, we fix a threshold $\alpha_{\omega} \in [0,1]$ and identify regions in which the vorticity is larger (or smaller) than $\alpha_{\omega}$ times the maximum (or minimum) vorticity (SM~\cite{SM}). This thresholding divides the sphere into patches of high absolute vorticity (Fig.~S1). The number of connected domains, given by the zeroth Betti number, counts the vortices in the system. For a fixed pair $(\kappa,R)$,  we denote the  vortex number at time~$t$ by $N_{\omega}(\kappa, R ; t)$. Although more sophisticated methods for vortex detection exist~\cite{jiang2005detection}, the thresholding criterion proved to be sufficiently robust for our analysis (Fig.~S2). To normalize vortex numbers across the parameter space, we fix a reference value $\kappa_* = 0.3/\Lambda$. With this, we can define a normalized Betti number as
\begin{equation}
\mrm{Betti}_\omega(\kappa, R) = \frac{\langle N_{\omega}(\kappa, R ; t) - N_{\omega}({\kappa_*}, R ; t) \rangle}{\langle N_{\omega}({\kappa_*}, R ; t) \rangle},
\end{equation}
where the time average $\lan \,\cdot\;\ran$ is taken after the initial relaxation period. Intuitively, large values of $\mrm{Betti}_\omega$ indicate many vortices of comparable circulation, whereas small values   suggest the presence of a few dominant eddies. The variation of $\mrm{Betti}_\omega$ in the $(\kappa, R)$- parameter plane  is color-coded in Fig.~\ref{fig:fig2}(a). In the anomalous turbulent A-phase, vortices of diameter $\approx\!\Lambda$ eventually cover the surface of the sphere, with stronger vortices forming chains of anti-ferromagnetic order [Fig.~\ref{fig:fig2}(d) top; Movie 4]. By contrast, in the T-phase characterized by broadband energy injection $\kappa>\Gl^{-1}$, smaller eddies merge to create a small number of larger vortices, as typical of  classical 2D turbulence~\cite{boffetta2012two} [Fig.~\ref{fig:fig2}(e) top; Movie 5].

\par
To obtain a more detailed geometric characterization of the turbulent A- and T-phases, we next consider the corresponding tension fields. Analogously to the case of vorticity above, we focus on regions where the local tension $\sigma(t,\bs x)$ is larger than the instantaneous global mean value. For each  connected component of the identified high-tension regions, we denote by $A$ its total area and by ${\partial}A$ its total boundary area in pixels. The ratio $\partial A/{A}$ is a measure of chain-like structures in the tension fields, a large value signaling  a highly branched structure, whereas smaller values indicate less branching. Denoting the instantaneous sum of the ratios ${\partial A}/A$ over all connected high-tension domains by $A_\sigma(\kappa, R ; t)$, a normalized branching index  can then  be defined by (SM~\cite{SM})
\begin{equation}
\mrm{Branch}_\sigma(\kappa, R) = \frac{\langle A_\sigma(\kappa, R ; t) - A_\sigma({\kappa_*}, R ; t) \rangle}{\langle A_\sigma({\kappa_*}, R ; t) \rangle},
\end{equation}
where the time average is again taken after the initial relaxation. As evident from the phase diagram in Fig.~\ref{fig:fig2}(b) and the corresponding tension fields in Fig.~\ref{fig:fig2}(d,e) and Movies 4,5, the geometric characterization  confirms the existence of an anomalous 
turbulent phase, in which vortices combine to form percolating dynamic networks with high-tension being localized along the edges [Fig.~\ref{fig:fig2}(d) bottom; Movie~4].
\par

\begin{figure}[t!]
\includegraphics[width=1\textwidth]{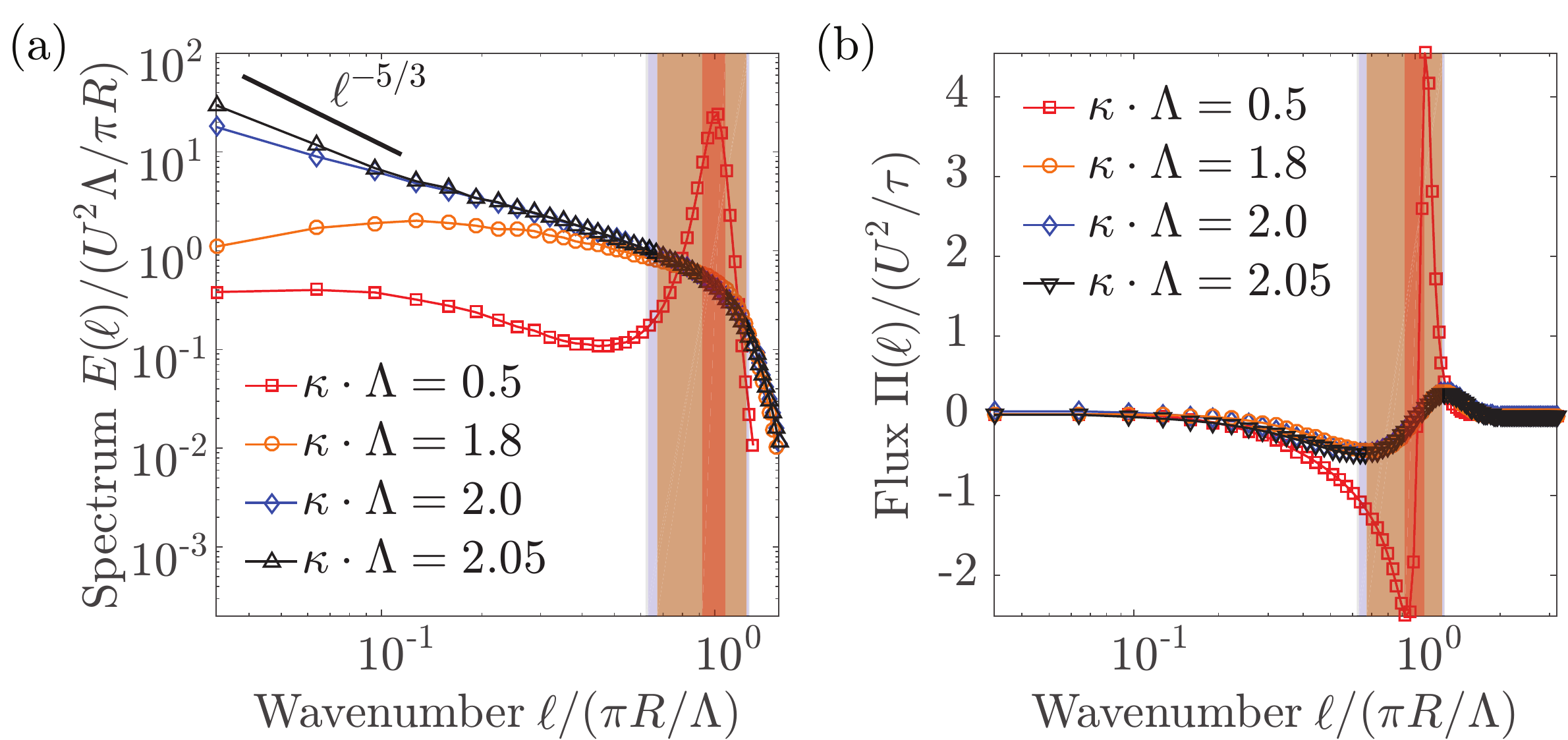}
\caption{Time-averaged energy spectra and fluxes indicate two qualitatively different types of upward energy transport.
(a)~For narrowband energy injection $\kappa \Gl < 1 $, the energy spectrum exhibits a peak corresponding to the dominant vortex size~$\Gl$ (red curve). 
For broadband injection $\kappa \Gl \sim 2  $, the spectra approach classical Kolmogorov scaling (blue and black curves). 
(b)~In all four examples, the fluxes confirm inverse energy transport, albeit with different origins.
For broadband energy injection  (blue and black curves), the upward energy flux to larger scales is due to vertex mergers [Fig.~\ref{fig:fig2}(e); Movie~5]. 
By contrast, for narrowband injection (red curve), a relatively stronger upward energy flux arises from the collective motion of vortex chains [Fig.~\ref{fig:fig2}(d); Movie~4]. 
The shaded regions indicate the energy injection ranges with colors matching those of the corresponding curves, respectively.  Simulation parameters: $R/\Lambda = 10$ for a unit sphere, $\tau = 11.7 \text{s}$, time step $5\cdot 10^{-4}\tau $, and total simulation time $500 \tau$. Spectra and fluxes were determined after relaxation, by averaging from $150\tau$ until the end of a simulation. For larger values~$\kappa\Gl\gg 1$, energy steadily accumulates at larger scales; in this case, the absence of a large-scale dissipative mechanism leads to a divergent total enstrophy and kinetic energy on the sphere.}
\label{fig:fig3}
\end{figure}

To compare the energy transport in the anomalous turbulent phase with classical 2D turbulence, we analyze the energy spectra and fluxes for the A- and T-phases. Expanding in spherical harmonics, $\psi = \sum_{m,\ell} \psi_{m\ell} Y_\ell^m$, the energy of mode $\ell$ is $E(\ell)= \sum_{\abs{m} \leq \ell} \ell(\ell+1)\abs{\psi_{m\ell}}^2$. The corresponding mean energy flux across $\ell$ in the statistically stationary state is obtained as (SM~\cite{SM})
\begin{equation}
\Pi(\ell) = -2 \sum_{\ell' \geq \ell} f[4-\ell'(\ell'+1)][2-\ell'(\ell'+1)] \langle E_{\ell'} \rangle,
\end{equation}
where $f$ is the polynomial defined in~Eq.~(\ref{eq:stress_tensor}). Figure~\ref{fig:fig3} shows the numerically obtained energy spectra $E(\ell)$ and fluxes $\Pi(\ell)$ for four active bandwidths $\kappa$. In all four cases, the kinetic energy produced in the injection range ($\ell \sim \pi R/\Lambda$) propagates to both large ($\ell< \pi R/\Lambda$) and small ($\ell > \pi R/\Lambda$) scales, as indicated by negative and positive values of $\Pi(\ell)$, respectively. Energy transfer to large scales is a prominent feature of classical 2D turbulence~\cite{kraichnan1967inertial,tang1978two,kraichnan1980two,boffetta2012two,sukoriansky2002universal} and our results show that it also occurs in active turbulence. However, the transfer mechanisms can be dramatically different, as already implied by the preceding analysis of the vorticity and tension fields. For broadband spectral forcing~$\kappa \Gl\gg 1$, the classical 2D turbulence picture of vortex mergers and energy condensation at large scales prevails [Fig.~\ref{fig:fig2}(e); Movie 3]. By contrast, for narrowband driving $\kappa\Gl \lesssim 1$, the upward energy transfer is realized through the coherent motion of high-tension vortex chains. Interestingly, only this anomalous type of inverse energy cascade appears to persist in 3D active bulk fluids~\cite{slomka2017spontaneous}, where it is sustained by spontaneous chiral symmetry-breaking~\cite{biferale2012inverse}.

\par

In summary, we have presented analytical and numerical solutions for generalized Navier-Stokes equations describing actively driven non-equilibrium flows on a sphere. Our calculations predict that spectrally localized active stresses can induce a novel turbulent phase, in which finite-size vortices self-organize into chain complexes of anti-ferromagnetic order   that percolate through the surface. The collective motion of these chain networks can enable a significant upward energy transport and may thus provide a basis for efficient fluid mixing in quasi-2D active and magnetohydrodynamic flows.

\begin{acknowledgments}
O.M. performed the numerical simulations and J.S. the analytical calculations.  O.M. and J.S. contributed equally. K.J.B, D.L. and  G.M.V. developed the numerical simulations. D.L. is supported by a Princeton Center for Theoretical Science Fellowship, and a Lyman Spitzer Jr. Fellowship. G.M.V. acknowledges support from The Australian Research Council project no. DE140101960. 
This work was supported by an Alfred P. Sloan Research Fellowship (J.D.) and a James S. McDonnell Foundation Complex Systems Scholar Award (J.D.). 
\end{acknowledgments}


%


\renewcommand{\thefigure}{S\arabic{figure}}
\setcounter{figure}{0}

\section*{Supplemental Material}

\section*{Vorticity-stream formulation}
Here, we derive the vorticity-stream function formulation~(2) of the Main Text from the momentum conservation~(1b) of the Main Text. We proceed as follows. We first rewrite the momentum equation explicitly in terms of the velocity vector field $v^a$. This involves the standard computation of the divergence of the rate of strain on curved surfaces as well as commuting covariant derivatives acting on rank-2 symmetric traceless tensors, which generates additional curvature terms. Once the equation for the contravariant vector $v^a$ is known, we lower the indices to find the corresponding equation satisfied by the covariant vector $v_a$. We then apply the Hodge decomposition~\cite{schwarz1995hodge} to $v_a$, which determines the unique (up to a constant) stream function $\psi$, and take the (surface) curl of the equation of motion for $v_a$ to find the equation for the vorticity function $\omega$. The final system takes the simple form of a higher order partial differential equation for two scalar functions $\psi$ and $\omega$ on a sphere, where the differential operators are the familiar Jacobian and spherical Laplacian.

\subsection{Divergence of rate of strain tensor}
To express the momentum conservation~(1b) of the Main Text in terms of the velocity field $v^a$, we must compute the divergence $\nabla_bT^{ab}$ of the stress tensor $T^{ab}$
\bse
\be
T^{ab}&=&f(\nabla^2)2S^{ab}, \\\
f(\nabla^2)&=&\Gamma_0-\Gamma_2\nabla^2+\Gamma_2\nabla^2\nabla^2,
\ee
\ese
where $\nabla^2=\nabla^c \nabla_c$ is the tensor Laplacian defined in terms of the covariant (Levi-Civita) derivative $\nabla_c$ and
\be
S^{ab}=\f{1}{2}(\nabla^a v^{b}+\nabla^b v^{a}),
\ee
is the rate of strain tensor for a stationary surface~\cite{scriven1960}. To this end, we first recall the standard calculation of the divergence of the rate of strain tensor $\nabla_a S^{ab}$, see for example~\cite{aris1989vectors}. By definition, the divergence is
\be
\label{e:rateofstrain_def}
2\nabla_a S^{ab}&=&\nabla_a\nabla^a v^b+\nabla_a\nabla^b v^a.
\ee
Before we simplify the last term by using the incompressibility condition $\nabla_a v^a=0$, we must compute the commutator $[\nabla_a,\nabla^b] v^a$. From the definition of the Riemann curvature tensor $R_{edac}$, we have
\be
\nabla_a\nabla_c v_d-\nabla_c\nabla_a v_d&=&-v^e R_{edac}.
\ee
For a sphere of radius $R$, the Riemann tensor is
\be
R_{edac}=K (g_{ea}g_{cd}-g_{ec}g_{ad}),
\ee
where $K=R^{-2}$ is the Gaussian curvature. In this special case, the commutator becomes
\be
\begin{aligned}
\nabla_a\nabla_c v_d-\nabla_c\nabla_a v_d &= -v^e K (g_{ea}g_{cd}-g_{ec}g_{ad}) \\
&=-K(v_ag_{cd}-v_cg_{ad}).
\end{aligned}
\ee
Contracting $a$ and $d$, and using the incompressibility $\nabla_a v^a=0$ condition we obtain
\be
\nabla^a\nabla_c v_a=Kv_c.
\ee
Substituting this for the last term in Eq.~(\ref{e:rateofstrain_def}), we recover the known result~\cite{aris1989vectors} that the divergence of the rate of strain tensor has the form
\be
\label{eq:div_rateofstrain}
2\nabla_a S^{ab}&=&\nabla_a\nabla^a v^b+Kv^b=(\nabla^2+K)v^b,
\ee
which implies that, on curved spaces, additional forces arise due to the curvature, beyond the viscous term proportional to the Laplacian.

\subsection{Computing $[\nabla_a,\nabla^2]$ on rank-two symmetric traceless tensors}
Now that we have the divergence of the rate of strain tensor~(\ref{eq:div_rateofstrain}), we can express the divergence of the stress tensor $\nabla_bT^{ab}$ in terms of $v^a$. To this end, we must know how to commute the tensor Laplacian $\nabla^2=\nabla^c\nabla_c$ with the operator $\nabla_a$. Since the velocity field is incompressible, the rate of strain tensor is traceless $S^a_a=0$ and so is any tensor Laplacian of it, $\nabla^2 S^a_a=0$ and $\nabla^2\nabla^2 S^a_a=0$. We thus have to compute the commutator $[\nabla_a,\nabla^2]$ on a symmetric and traceless rank-2 tensor. The following calculation is very similar to that presented in Section 3 in~\cite{delay2007tt}. For a rank-three covariant tensor $H_{cde}$ we have
\be
\begin{aligned}
\label{eq:com_rank3tensors}
[\nabla_a,\nabla_b]H_{cde}=
&-R^{f}{}_{cab}H_{fde}-R^f{}_{dab}H_{cfe}\\
&-R^f{}_{eab}H_{cdf},
\end{aligned}
\ee
Contracting $a$ with $d$ and $b$ with $c$ yields
\be
\begin{aligned}
\label{eq:com_rank3tensors_contract}
[\nabla^a,\nabla^b]H_{bae}= 
&-R^{f}{}^{b}{}^{a}{}_bH_{fae}-R^f{}^{a}{}_{a}{}^{b}H_{bfe} \\
&-R^f{}_{e}{}^{ab}H_{baf}.
\end{aligned}
\ee
The first two terms on the right hand side cancel out
\be
\begin{aligned}
\label{eq:com_rank3tensors_cancellation}
R^{f}{}^{b}{}^{a}{}_bH_{fae}+&
R^f{}^{a}{}_{a}{}^{b}H_{bfe}
\\=&
(R^{f}{}^{b}{}^{a}{}_b+R^a{}^{b}{}_{b}{}^{f})H_{fae}
\\ =&
(R_{f}{}^{b}{}_{a}{}_b+R_a{}^{b}{}_{b}{}_{f})H^{fa}{}_{e}
\\ =&
(R_{f}{}_{h}{}_{a}{}_b+R_a{}_{b}{}_{h}{}_{f})g^{bh}H^{fa}{}_{e}
\\ =&0
\end{aligned}
\ee
since $R_{f}{}_{h}{}_{a}{}_b=R_{h}{}_{f}{}_{a}{}_b=-R_{a}{}_{b}{}_{h}{}_f$, which reduces~(\ref{eq:com_rank3tensors_contract}) to
\be
\label{eq:com_piece1}
[\nabla^a,\nabla^b]H_{bae}=
R^f{}_{e}{}^{ab}H_{abf}.
\ee

Similarly, for rank-two covariant tensors, we have
\be
\label{eq:com_rank2tensors}
[\nabla_a,\nabla_b]H_{cd}=-R^e{}_{cab}H_{ed}-R^e{}_{dab}H_{ce}.
\ee
Contracting $a$ and $c$ gives
\be
(\nabla^a\nabla_b-\nabla_b\nabla^a)H_{ad}=R^e{}_{b}H_{ed}-R^e{}_{d}{}^a{}_bH_{ae},
\ee
where $R^e{}_b$ is the Ricci tensor, and further contraction with $\nabla^b$ yields
\be
\begin{aligned}
(\nabla^b\nabla^a\nabla_b-\nabla^b\nabla_b\nabla^a)H_{ad}=&
\nabla^b(R^e{}_{b}H_{ed})\\
&-\nabla^b(R^e{}_{d}{}^a{}_bH_{ae}).
\quad
\end{aligned}
\ee
We restrict to the specific case of a sphere, for which the Riemann tensor reads
\be
\begin{aligned}
R_{abcd}&=K(g_{ac}g_{db}-g_{ad}g_{cb}), \\
R^a_{\;bcd}&=K(\delta^a_c g_{db}-\delta^a_d g_{cb}),
\end{aligned}
\ee
where $K$ is the Gaussian curvature. The Ricci tensor is $R_{ab}=Kg_{ab}$. Since $K$ is constant and we are working with the Levi-Civita connection, we have
\be
\begin{aligned}
(\nabla^b\nabla^a\nabla_b-&\nabla^b\nabla_b\nabla^a)H_{ad}\\
&=
R^e{}_{b}\nabla^bH_{ed} -R^e{}_{d}{}^a{}_b\nabla^bH_{ae} \\
&=
K\nabla^eH_{ed} -R^e{}_{d}{}^a{}_b\nabla^bH_{ae}.
\end{aligned}
\ee
The last term is
\be
R^e{}_{d}{}^a{}_b\nabla^bH_{ae}=K\nabla_dH^a_a-K\nabla^eH_{de}.
\ee
Since our symmetric tensor is traceless, we have
\be
\label{eq:com_piece2}
(\nabla^b\nabla^a\nabla_b-\nabla^b\nabla_b\nabla^a)H_{ad}
&=&
2K \nabla^eH_{de}.
\ee
 We now combine Eq.~(\ref{eq:com_piece1}) with Eq.~(\ref{eq:com_piece2}) by setting $H_{bae}=\nabla_b H_{ae}$. For this choice, Eq.~(\ref{eq:com_piece1}) gives
\be
\nabla^a\nabla^b\nabla_bH_{ae}=\nabla^b\nabla^a\nabla_b H_{ae}+
R^f{}_{e}{}^{ab}\nabla_a H_{bf}.
\ee
Using (\ref{eq:com_piece2}) to replace the first term on the right hand side gives
\be
\begin{aligned}
\nabla^a\nabla^b\nabla_bH_{ae}= &
\nabla^b\nabla_b\nabla^a H_{ae}
+2K \nabla^fH_{ef}\\
&+R^f{}_{e}{}^{ab}\nabla_a H_{bf}.
\end{aligned}
\ee
We rewrite the last term above as
\be
\begin{aligned}
R^f{}_{e}{}^{ab}\nabla_a H_{bf}&= K(g^{fa}\delta^b_e-g^{fb}\delta^a_e)\nabla_a H_{bf} \\
&=
K\nabla^f H_{ef}-K\nabla_e H_{f}^f=K\nabla^f H_{ef}.
\end{aligned}
\ee
Finally, we obtain the following expression
\be
\begin{aligned}
\label{eq:commutator_rank2}
\nabla_a\nabla^b\nabla_bH^{ae}&=
\nabla^b\nabla_b\nabla_a H^{ae}
+3K \nabla_a H^{ae}\\
&=
(\nabla^2
+3K) \nabla_a H^{ae},
\end{aligned}
\ee
which allows us to commute the tensor Laplacian with the divergence operator for any symmetric and traceless tensor $H_{ae}$ on a sphere with the Gaussian curvature $K$.

\subsection{Divergence of stress tensor and equation for $v^a$}
To calculate the divergence of the stress tensor, we combine the results of the two previous subsections. We obtain
\be
\begin{aligned}
\nabla_bT^{ab}&=\nabla_b f(\nabla^2)2S^{ab}=
f(\nabla^2+3K)2\nabla_b S^{ab} \\
&=
f(\nabla^2+3K)(\nabla^2+K)v^b,
\end{aligned}
\ee
where we used Eq.~(\ref{eq:commutator_rank2}) in the first line, remembering that $S^a_a=\nabla^2 S^a_a=\nabla^2\nabla^2S^a_a=0$, and Eq.~(\ref{eq:div_rateofstrain}) in the second line.
We can now express the momentum conservation equation (1b) of the Main Text solely in terms of the velocity vector field $v^a$
\be
\begin{aligned}
\label{eq:velocity_field}
\p_t v^a+v^b \nabla_bv^a = & f(\nabla^2+3K)(\nabla^2 +K)v^a \\
&+\nabla^a \sigma.
\end{aligned}
\ee

\subsection{Coordinate-free equation for $v_a$}
We lower the indices in~(\ref{eq:velocity_field}) to obtain the corresponding equation for the covariant vector (one-form) $v_a$
\be
\begin{aligned}
\p_t v_a+v^b \nabla_bv_a =&f(\nabla^2+3K)(\nabla^2 +K)v_a \\
&+\nabla_a \sigma.
\end{aligned}
\ee
Since we are ultimately interested in applying the Hodge decomposition to the one-form $v_a$, we first replace the tensor Laplacian $\nabla^2=\nabla^c\nabla_c$ with the Hodge Laplacian $\Delta_H=\delta d+ d\delta$, where $d$ and $\delta$ are the differential and codifferential operators. The Weitzenbock identity~\cite{besse2007einstein} for one-forms reads
\be
\Delta_H v_a=-\nabla^c\nabla_c v_a+R_{ab}v^b,
\ee
where $R_{ab}$ is the Ricci tensor. For a sphere, we have $R_{ab}=Kg_{ab}$, and the above identity reduces to
\be
\Delta_H v_a=-\nabla^c\nabla_c v_a+Kv_a.
\ee
In terms of the Hodge Laplacian, the equations of motion read
\be
\label{e:eom_Hodge_Laplacian}
\begin{aligned}
\nabla^av_a=&0, \\
\p_t v_a+v^b \nabla_bv_a= &f(-\Delta_H+4K)(-\Delta_H +2K)v_a \qquad\\
&+\nabla_a \sigma
\end{aligned}
\ee
To simplify the subsequent calculations, we rewrite the above equations in the coordinate-free form. Denote by $\bs v$ the contravariant field $v^a$ and by $v$ the corresponding covariant one-form $v_a$. In the new notation~(\ref{e:eom_Hodge_Laplacian}) reads
\be
\begin{aligned}
\delta v&=0, \\
\p_t v+\nabla_{\bs v}v&=d\sigma+f(-\Delta_H+4K)(-\Delta_H +2K)v,
\quad
\end{aligned}
\ee
where again, $d$ and $\delta$ denote the differential and codifferential operators. It is useful to apply the following identity relating the Lie and covariant derivatives of one forms~\cite{arnold1999topological}
\be
L_{\bs v}v=\nabla_{\bs v}v+\f{1}{2}d \bs v^2,
\ee
where $\bs v^2/2=v^av_a/2$ is the kinetic energy density.  In terms of the Lie derivative, the equation of motion reads
\bse
\label{eq:EOMcoordinatefree}
\begin{align}
\label{eq:EOMcoordinatefree_a}
\delta v=&0,\\
\label{eq:EOMcoordinatefree_b}
\begin{split}
\p_t v+L_{\bs v}v= & f(-\Delta_H+4K)(-\Delta_H +2K)v\\
&+ d(\sigma+\f{1}{2}\bs v^2).
\end{split}
\end{align}
\ese

\subsection{Hodge decomposition and vorticity-stream function formulation}

Since $v$ is co-exact ($\bs v$ is divergence-free), we can use the Hodge decomposition~\cite{schwarz1995hodge} to write $v=\delta \tilde\psi$, for some two-form (pseudoscalar) $\tilde\psi$ that we will later identify with the stream function. Introducing $\tilde\psi$ automatically satisfies Eq.~(\ref{eq:EOMcoordinatefree_a}). We take the differential $d$ of Eq.~(\ref{eq:EOMcoordinatefree_b}) to derive the vorticity equation. The great advantage of the above coordinate-free representation is that the differential commutes both with the Lie derivative and Hodge Laplacian. We get
\be
\p_t \tilde\omega+L_{\bs v}\tilde\omega&=&f(-\Delta_H+4K)(-\Delta_H +2K)\tilde\omega,
\ee
where we introduced the vorticity pseudoscalar $\tilde\omega=dv=d\delta \tilde\psi=\Delta_H \tilde\psi$. Thus, the equations of motion become
\be
\begin{aligned}
\Delta_H \psi&=\tilde\omega, \\
\p_t \tilde\omega+L_{\bs v}\tilde\omega&=f(-\Delta_H+4K)(-\Delta_H +2K)\tilde\omega.
\end{aligned}
\ee
Above, $\tilde\psi$ and $\tilde\omega$ are both pseudoscalars (two-forms in 2D). We now apply the Hodge star $\ast$, which commutes with the Hodge Laplacian, to find equations for the scalars $\omega=\ast \tilde\omega$ and $\psi=\ast \tilde\psi$
\be
\begin{aligned}
\Delta_H \psi&=\omega, \\
\p_t \omega+\ast L_{\bs v}\tilde\omega&=f(-\Delta_H+4K)(-\Delta_H +2K)\omega.
\end{aligned}
\ee
Since on scalars the Hodge Laplacian equals the negative of the familiar Laplace-Beltrami operator, $\Delta_H=-\Delta$, we finally arrive at
\be
\begin{aligned}
\label{eq:vorticity_stream_fun_coordinate_free}
\Delta \psi&=-\omega, \\
\p_t \omega+\ast L_{\bs v}\tilde\omega&=f(\Delta+4K)(\Delta +2K)\omega.
\end{aligned}
\ee

\subsection{Spherical coordinates}
In this section, we explicitly write Eqs.~(\ref{eq:vorticity_stream_fun_coordinate_free}) in spherical coordinates $(\theta,\phi)$. The metric and its inverse are $g_{ij}=R^2\textnormal{diag}(1,\sin^2\theta)$ and $g^{ij}=R^{-2}\textnormal{diag}(1,1/\sin^2\theta)$ and the determinant volume prefactor is $\sqrt{|g|}=R^2\sin\theta$. 

Let the pseudoscalar stream function be
\be
\begin{aligned}
\tilde\psi&=\psi(\theta,\phi)\sqrt{|g|} d\theta\wedge d\phi\\
&=\psi(\theta,\phi)R^2\sin\theta d\theta\wedge d\phi,
\end{aligned}
\ee
where $\psi=\ast\tilde\psi$ is the stream function on the sphere. We now compute the velocity field $v=\delta\tilde\psi$.
The codifferential for a 2D Riemannian manifold is $\delta=-\ast d \ast$, and since $\psi=\ast\tilde\psi$, we obtain
\be
d\ast\tilde\psi=\p_\theta \psi d\theta+\p_\phi \psi d\phi.
\ee
Applying the Hodge star yields
\be
\begin{aligned}
\ast d\ast\tilde\psi &=(d\ast\tilde\psi)^i \sqrt{|g|}\epsilon_{ij}dx^j \\
&=-\f{1}{\sin\theta}\p_\phi \psi d\theta+\p_\theta \psi \sin\theta d\phi,
\end{aligned}
\ee
and the velocity one-form becomes
\be
v=\delta\tilde\psi=\f{1}{\sin\theta}\p_\phi \psi d\theta-\p_\theta \psi \sin\theta d\phi.
\ee
By raising the indices, we obtain the corresponding velocity vector field
\be
\bs v=\f{1}{R^2}\f{1}{\sin\theta}\p_\phi \psi \p_\theta-\f{1}{R^2}\f{1}{\sin\theta}\p_\theta \psi \p_\phi.
\ee
In terms of the usual unit vectors $\hat{\bs\theta}$ and $\hat{\bs\phi}$, this is $\bs v=\f{1}{R}\f{1}{\sin\theta}\p_\phi \psi \hat{\bs\theta}-\f{1}{R}\p_\theta \psi\hat{\bs\phi}$. We now compute the vorticity pseudoscalar
\be
\begin{aligned}
\tilde\omega=dv&=-
\biggl[
\f{1}{\sin\theta}\p_{\phi\phi}\psi+\p_\theta(\sin\theta\p_\theta\psi)
\biggr]d\theta\wedge d\phi \\
&=
\omega(\theta,\phi)R^2\sin\theta d\theta\wedge d\phi,
\end{aligned}
\ee
where 
\be
\begin{aligned}
\omega(\theta,\phi)=&-\f{1}{R^2}\biggl[\f{1}{\sin\theta}\p_\theta(\sin\theta\p_\theta)\\
&\qquad\quad+\f{1}{\sin^2\theta}\p_{\phi\phi}\biggr]\psi(\theta,\phi),
\end{aligned}
\ee
and we recover $\Delta \psi=-\omega$ as required, since $\omega=\ast \tilde\omega=\omega(\theta,\phi)$.
We finally compute the Lie derivative
\be
L_{\bs v}\tilde\omega=d i_{\bs v}\tilde\omega.
\ee
We start with
\be
\begin{aligned}
i_{\bs v}\tilde\omega &=\omega(\theta,\phi)R^2\sin\theta(v^\theta d\phi -v^\phi d\theta)\\
&=\omega(\theta,\phi)\p_\theta\psi d\theta+\omega(\theta,\phi)\p_\phi\psi d\phi,
\end{aligned}
\ee
then
\be
\begin{aligned}
L_{\bs v}\tilde\omega&=
\Big[\p_\theta(\omega\p_\phi\psi) -\p_\phi(\omega \p_\theta\psi) \Big] d\theta\wedge d\phi \\
&=
\Big(\p_\theta\omega\p_\phi\psi -\p_\phi\omega \p_\theta\psi \Big) d\theta\wedge d\phi.
\end{aligned}
\ee
Finally,
\be
\ast L_{\bs v}\tilde\omega&=&\f{1}{R^2}\f{1}{\sin\theta}(\p_\theta\omega\p_\phi\psi -\p_\phi\omega \p_\theta\psi ).
\ee
To sum up, in spherical coordinates, the equations of motion [Eq.~(2) in the Main Text] read
\bse
\label{e:EOMsphericalcoords}
\be
\label{e:EOMsphericalcoordsA}
\Delta \psi&=&-\omega, 
\\
\label{e:EOMsphericalcoordsB}
\p_t \omega &=& \f{1}{R^2}\f{1}{\sin\theta}(\p_\theta\omega\p_\phi\psi -\p_\phi\omega \p_\theta\psi ) \\
&&+f(\Delta+4K)(\Delta +2K)\omega, \nonumber
\ee
\ese
where $\omega$ and $\psi$ are scalars (not pseudoscalars) on a sphere and $\Delta$ is the usual Laplacian in spherical coordinates.

\subsection{Expansion in spin-weighted spherical harmonics}

For numerical integration within Dedalus~\cite{dedalus2017}, each vector and tensor is expanded in terms of a basis of coherent spin weight: $\mathbf{e}_{\pm} = (\mathbf{e}_{\theta} \pm i \mathbf{e}_{\phi})/\sqrt{2}$. A unitary matrix transforms between the spin basis and the coordinate basis. The tensor product of unit vectors of coherent spin weight adds their individual spin, i.e. the rank-4 tensor basis element $\mathbf{e}_{+} \mathbf{e}_{-} \mathbf{e}_{+}\mathbf{e}_{+}$ carries spin-weight $+1-1+1+1=2$. We expand the  components of a tensor in terms of spin-weighted spherical harmonics deepening on the spin-weight of their basis vectors. The surface tension, $\sigma$,  is a pure spin-0 field. The velocity, $v$, is a sum of $\pm1$ components. Higher-order tensors comprise a range of spin-weights; e.g., $v \otimes v$ contains components with spin-$\pm2$ and spin-0.  

For example, with an rank-$r$ tensor, 
\begin{eqnarray}
\mathrm{T} =  
\sum_{\sigma_{i} = \pm1} \sum_{m=-L}^{L} \sum_{\ell = \ell_{0}}^{L} T_{\ell,m}^{\sigma_{1},\ldots \sigma_{r}} Y_{\ell}^{m,s} (\theta,\phi)\mathbf{e}_{\sigma_{1}} \ldots \mathbf{e}_{\sigma_{r}}
\quad
\end{eqnarray}
where $s = \sigma_{1} + \ldots + \sigma_{r}$, $\ell_{0} = \max(|m|,|s|)$, and $ T_{\ell,m}^{\sigma_{1},\ldots \sigma_{r}}$ represent an array of spectral coefficients. The intrinsic gradient operator on the two-sphere acts in a coherent way with regard to spin. 
\begin{eqnarray}
&&\notag
\nabla \left(Y_{\ell}^{m,s}  \mathbf{e}_{\sigma_{1}} \ldots \mathbf{e}_{\sigma_{r}}   \right) 
\\
&&= \left( k_{\ell,s}^{+}Y_{\ell}^{m,s+1}  \mathbf{e}_{+}  \ + \ k_{\ell,s}^{-}Y_{\ell}^{m,s-1}  \mathbf{e}_{-} \right) \mathbf{e}_{\sigma_{1}} \ldots \mathbf{e}_{\sigma_{r}}
\qquad
\end{eqnarray}
where
\begin{eqnarray}
k^{\mu}_{\ell,s} \ =  \ -\mu \sqrt{\frac{(\ell - \mu s)(\ell + \mu s+1)}{2}}.
\end{eqnarray}
This means that the spectral coefficients act in a particularly simple way under differentiation. 
\begin{eqnarray}
\nabla \mathrm{T} \quad  \longleftrightarrow \quad k_{\ell,s}^{\pm}\, T_{\ell,m}^{\sigma_{1},\ldots \sigma_{r}}
\end{eqnarray}
The spin-weighed basis renders computations in the sphere almost identical to Fourier series from an algorithmic perspective. Said another way: the gradient of a traditional spherical harmonic function is not a series of traditional spherical harmonic functions. But it is a very small number of other kinds of functions. This is philosophically the same a saying that the derivative of a cosine function is not naturally a series of cosines functions. But it is a very simple expression in terms of sine functions, and vice versa. In fact, sine and cosine functions {\it are} the spherical harmonic basis for the one-dimensional sphere (also known as the circle); so it's more than just a convenient analogy, the same underlying structure is at play in both cases. 
\par
The spin-weighted spherical harmonic function are each orthonormal under integration on the unit sphere. We therefore use Gauss quadrature to transform from field on a Legendre quadrature grid to the spectral coefficients. Linear operations happen in spectral space, nonlinear multiplications happen locally on the grid. 
\par
With the above definitions, we can compute the {\it intrinsic} Laplacian in two-dimensions. Acting on an individual spin component, 
\begin{eqnarray}
\nabla \cdot \nabla \mathrm{T} \;  \longleftrightarrow \; \left( k^{-}_{\ell,s+1} k^{+}_{\ell,s} + k^{+}_{\ell,s-1} k^{-}_{\l,s}\right) T_{\ell,m}^{\sigma_{1},\ldots \sigma_{r}},
\quad
 \label{rough Laplacian}
\end{eqnarray}
where 
\begin{eqnarray}
 k^{-}_{\ell,s+1} k^{+}_{\ell,s} + k^{+}_{\ell,s-1} k^{-}_{\l,s} \ = \ - \ell (\ell+1) + s^{2}.
\end{eqnarray}
Equation~\eqref{rough Laplacian} gives a slightly different formula than would result from taking the three-dimensional Laplacian and restricting the result to the surface of the unit 2-sphere. In this case, additional terms result from contracting in the third dimension. Equation \eqref{rough Laplacian} defines what is often called the \textit{rough Laplacian}, \textit{connection Laplacian}, or \textit{intrinsic Laplacian}. In curved geometry, it is possible to define several linear, second-order,  elliptic differential operators with a reasonable claim to the title of Laplacian. The Weitzenb\"{o}ck identity ensures that any two such Laplacians differ by a scalar curvature term at most; i.e., a term with no derivatives.  In the case of the restricted three-dimensional Laplacian, 
\begin{eqnarray}
\label{Other-Laplacian}
\notag
 \nabla_{3D}\cdot \nabla_{3D} \mathrm{T}  \big|_{S^{2}}  \quad  \longleftrightarrow \quad  -\left(\ell (\ell+1) - s^{2} + r \right)T_{\ell,m}^{\sigma_{1},\ldots \sigma_{r}},
\end{eqnarray}
where $r $ gives the tensor rank of $\mathrm{T} $. In the case of simple vectors, $\mathbf{e}_{\pm}$, $s^{2} = r = 1$. In applications, the most appropriate Laplacian depends on the details of the underlying physics. 


\section{Nondimensionalization and exact solutions}
\subsection{Nondimensionalization}
Before we construct exact stationary solutions of~(\ref{e:EOMsphericalcoords}), we first nondimensionalize the equation by introducing the length scale $R$ and the time scale $T$, and set
\be
\psi \to \f{R^2}{T}\psi, \quad \omega\to \f{1}{T}\omega.
\ee
For a sphere of radius $R$, the Gaussian curvature is \mbox{$K=R^{-2}$}. The equations of motion~(\ref{e:EOMsphericalcoords}) become
\be
\begin{aligned}
\Delta \psi=&-\omega, \\
\p_t \omega =&
f(\Delta+4)(\Delta +2)\omega \\
&- \f{1}{\sin\theta}(\p_\theta\omega\p_\phi\psi -\p_\phi\omega \p_\theta\psi ),
\end{aligned}
\ee
where $\Delta=(\sin\theta)^{-1}\p_\theta(\sin\theta\p_\theta)+(\sin\theta)^{-2}\p_{\phi\phi}$ is the usual Laplacian on the unit sphere and
\be
\begin{aligned}
f(\Delta+4)(\Delta +2)\omega=
&\,[1-\gamma_2(\Delta+4)+\gamma_4 (\Delta+4)^2]
\qquad\\
&\times\Gamma_0\f{T}{R^2}(\Delta +2)\omega,
\end{aligned}
\ee
where $\gamma_2=\Gamma_2/(\Gamma_0 R^2)$ and $\gamma_4=\Gamma_4/(\Gamma_0 R^4)$.
To summarize, the nondimensionalized equations are
\be
\label{e:EOMsphericalcoordsNondim}
\begin{aligned}
\Delta \psi&=-\omega, \\
\p_t \omega &= -  \f{1}{\sin\theta}(\p_\theta\omega\p_\phi\psi -\p_\phi\omega \p_\theta\psi ) \\
&\quad+[1-\gamma_2(\Delta+4)+\gamma_4 (\Delta+4)^2]\, (\Delta +2)\omega ,
\end{aligned}
\ee
where we set the time scale to $T=R^2/\Gamma_0$.

\subsection{Exact stationary solutions}
We start the construction of exact stationary solutions of~(\ref{e:EOMsphericalcoordsNondim}) by first noting that the spherical harmonics $Y^m_{\ell} (\theta,\phi)$ are eigenstates of the Laplace operator $\Delta=(\sin\theta)^{-1}\p_\theta(\sin\theta\p_\theta)+(\sin\theta)^{-2}\p_{\phi\phi}$
\be
\Delta Y^m_\ell=-\ell(\ell+1)Y^m_{\ell}.
\ee
Taking the linear combination $\psi=\sum_{m}a_m Y^m_{\ell}$, where $-\ell \leq m \leq \ell$ and $a_m$ are arbitrary real numbers, annihilates the nonlinear term in Eq.~(\ref{e:EOMsphericalcoordsB}) because $\omega=\ell(\ell+1)\psi$ by Eq.~(\ref{e:EOMsphericalcoordsA}). This reduces~(\ref{e:EOMsphericalcoordsB}) to the polynomial equation
\be
\label{e:exact_sol_root}
\begin{aligned}
0=&\Big\{1+\gamma_2\big[\ell(\ell+1)-4\big]+\gamma_4 \big[\ell(\ell+1)-4\big]^2\Big\}
\quad\\
&\times [\ell(\ell+1)-2].
\end{aligned}
\ee
The index $l$ is a non-negative integer; if it coincides with a positive root of the above polynomial, we obtain an exact stationary solution. There are two possibilities: either
\be
1+\gamma_2\big[\ell(\ell+1)-4\big]+\gamma_4 \big[\ell(\ell+1)-4\big]^2=0
\ee
or
\be
\ell(\ell+1)-2=0.
\ee
The first possibility is a direct consequence of the higher-order nature of Eqs~(\ref{e:EOMsphericalcoordsNondim}). In this case, non-trivial roots exist when $\gamma_2<0$, which introduces linearly unstable modes. Example of solutions of this type are shown in Fig.~1 of the Main Text. The second possibility gives $\ell=1$ as the only admissible solution and arises even for the classical Navier-Stokes equations. Superposition of $\ell=1$ spherical harmonics corresponds to flow patterns representing rigid rotation of the whole sphere with rotation rate and rotation axis specified by the three constants $\{a_{-1},a_{0},a_1\}$. We stress that this second possibility arises only when one derives the equations of motion from the Cauchy equations (Eq.~1 in the Main Text) instead of starting with the equations for the velocity field in the flat space and promoting the corresponding differential operators to covariant ones, implying that the latter approach is incorrect.

\section{Energy spectrum and energy flux}

\subsection{Energy spectrum}
For flows on periodic domains, the energy spectrum is typically defined by expanding energy using the Fourier series. For flows on a sphere, the most natural analogue is obtained by the  spherical harmonics basis.
The kinetic energy density function is $e=v_a v^a/2$. Integrating over the sphere surface gives the total kinetic energy
\be
E=\int_S e d\Omega=\f{1}{2}\int_S v_i v^i d\Omega,
\ee
where $d\Omega$ is the area element. In coordinate free notation
\be
\int_S v_i v^i d\Omega=\int_S v\wedge \ast v=\langle v,v \rangle,
\ee
where $\langle v,v \rangle$ is the inner product of one-forms. We use the Green's formula~\cite{schwarz1995hodge} (integration by parts) to get
\be
\begin{aligned}
2E&=\langle v,v \rangle =\langle \delta \tilde\psi ,\delta \tilde\psi \rangle=
\langle  \tilde\psi ,d\delta \tilde\psi \rangle=
\langle  \tilde\psi ,\Delta_H\tilde \psi\rangle \\
&
= \langle  \tilde\psi ,\tilde\omega \rangle
=\int_S \tilde \psi\wedge \ast\tilde\omega=
\int_S \tilde \psi\ \omega
=\int_S \psi\omega d\Omega.
\qquad
\end{aligned}
\ee
Since $\Delta\psi=-\omega$, we obtain for the total energy
\be
\begin{aligned}
E&=\f{1}{2}\int_S \psi\omega d\Omega=\f{1}{2}\int_S \psi\omega d\Omega\\
&=
-\f{1}{2}\int_S \psi\Delta\psi d\Omega.
\end{aligned}
\ee
Expanding $\psi$ in the spherical harmonics basis
\be
\psi=\sum_{m,\ell}\psi_{m\ell}Y^m_{\ell},
\ee
and applying the orthogonality condition
\be
\int_S Y^m_{\ell} Y^{m'}_{\ell'} d\Omega=\delta_{\ell \ell'}\delta_{mm'},
\ee
yields
\be
E=\f{1}{2}\sum_{m,\ell}\ell(\ell+1)|\psi_{m\ell}|^2=\f{1}{2}\sum_{\ell} E_{\ell},
\ee
where the energy in mode $l$ is given by 
\be
E_{\ell}=\sum_{|m|\leq \ell}\ell(\ell+1)|\psi_{m\ell}|^2. 
\ee

\subsection{Energy flux}
We derive the expression for the energy flux $\Pi(\ell)$ across the wavenumber $\ell$ shown in Fig.~3 of the Main Text.
\par
Denote by $\mathcal{S}$ the spherical harmonics transform, that maps a function defined on a sphere to its coefficients in the spherical harmonics basis. Apply $\mathcal{S}$ to the equations of motion to get
\bse
\label{eq:SHTeom}
\be
\label{eq:SHTeomA}
\omega_{m\ell} &=& \ell(\ell+1)\psi_{m\ell},
\ee
\be
\label{eq:SHTeomB}
\p_t \omega_{m\ell}&=& -\mathcal{S}_{m\ell}\Big\{\f{1}{\sin\theta}(\p_\theta\omega\p_\phi\psi -\p_\phi\omega \p_\theta\psi )\Big\}\\
&&+ \mathcal{F}(\ell)\omega_{m\ell}, \nonumber
\ee
\ese
where
\be
\begin{aligned}
\mathcal{F}(\ell)=&
\Big\{1-\gamma_2[4-\ell(\ell+1)]+\gamma_4 [4-\ell(\ell+1)]^2\Big\}\quad\\
&\times\Big[2-\ell(\ell+1)\Big].
\end{aligned}
\ee
Since the energy spectrum is given by $E_{\ell}=\sum_{|m|\leq \ell}\ell(\ell+1)|\psi_{m\ell}|^2$, we use~(\ref{eq:SHTeomA}) to substitute for $\omega_{m\ell}$ in~(\ref{eq:SHTeomB}) to find the evolution equation for $\psi_{m\ell}$
\be
\begin{aligned}
\ell(\ell+1)\p_t \psi_{m\ell} 
=& -\mathcal{S}_{m\ell}\Big\{\f{1}{\sin\theta}(\p_\theta\omega\p_\phi\psi -\p_\phi\omega \p_\theta\psi )\Big\} \qquad\\
&+\mathcal{F}(\ell)\ell(\ell+1)\psi_{m\ell}.
\end{aligned}
\ee
We multiply both sides by $\psi_{m\ell}^*$ and add the resulting equation to its complex conjugate to get
\be
\begin{aligned}
\ell(\ell+1)&\p_t |\psi_{m\ell}|^2\\
=&\,2\mathcal{F}(\ell)\ell(\ell+1)|\psi_{m\ell}|^2\\
&-\Big(\psi_{m\ell}^*\mathcal{S}_{m\ell}\Big\{\f{1}{\sin\theta}(\p_\theta\omega\p_\phi\psi -\p_\phi\omega \p_\theta\psi )\Big\}+\tn{c.c.}\Big).
\end{aligned}
\ee
Suming over $|m|\leq \ell$ and rearranging yields the evolution equation for $E_{\ell}$
\be
\begin{aligned}
\p_t E_{\ell}-2\mathcal{F}(\ell)E_{\ell} = -\sum_{|m|\leq \ell}\Big(\psi_{m\ell}^* &\mathcal{S}_{m\ell}\Big\{\f{1}{\sin\theta}(\p_\theta\omega\p_\phi\psi \\
&-\p_\phi\omega \p_\theta\psi )\Big\}+\tn{c.c.}\Big).\\
\end{aligned}
\ee

Since for the Euler equations on a sphere $\mathcal{F}(\ell)=0$ holds, we identify the right hand side above as the nonlinear energy transfer
into mode $\ell$.  In the statistically stationary state and after time averaging we expect $\langle \p_t E_{\ell} \rangle=0$ and hence
\be
\begin{aligned}
-2\langle\mathcal{F}(\ell)E_{\ell}\rangle 
= -\biggl\langle
\sum_{|m|\leq \ell}\Big(&\psi_{m\ell}^*\mathcal{S}_{m\ell}\Big\{\f{1}{\sin\theta}(\p_\theta\omega\p_\phi\psi 
\qquad\\
&-\p_\phi\omega \p_\theta\psi )\Big\}+\tn{c.c.}\Big)
\biggr\rangle.
\end{aligned}
\ee
The energy flux across $\ell$ is then finally given by
\be
\Pi(\ell)=-2\sum_{\ell'\geq \ell}\langle\mathcal{F}(\ell')E_{\ell'}\rangle.
\ee
In Eq.~(6) of the Main Text, we express the prefactor  $\mathcal{F}(\ell)$ in terms of the polynomial $f$ defined in~Eq.~1(c) of the Main Text as
\be
\mathcal{F}(\ell)= f\bigl(4-\ell(\ell+1)\bigr)\;[2-\ell(\ell+1)].
\ee

\begin{figure*}[t!]
\includegraphics[width=0.9\textwidth]{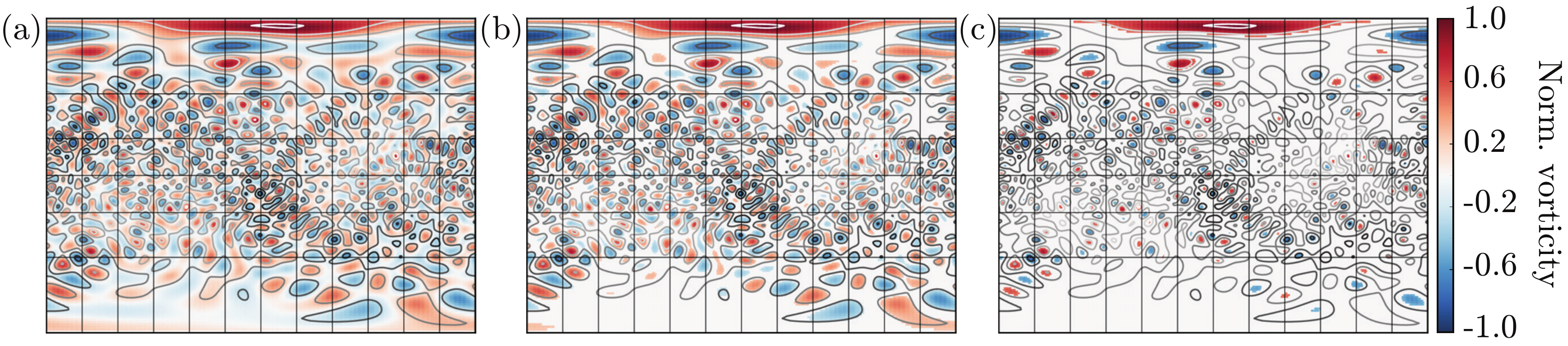}
\caption{Vortex detection scheme. Miller cylindrical projection of the sphere showing the unprocessed normalized vorticity field (a), and the thresholded vorticity field with values $\omega(x,t)\in[\alpha_{\omega}\min\limits_{x\in S^2}  \omega(x,t), \alpha_{\omega} \max\limits_{x\in S^2}]$ removed for (b) $\alpha_{\omega} = 0.25$, and (c) $\alpha_{\omega} = 0.5$. The chain-like branched structures in the vorticity field remain preserved after thresholding.
\label{fig:SM_detection}} 
\end{figure*}

\section{Expansion coefficients for exact solutions}
The spectrum of the instantaneous snapshot in Fig~2(c) shows a peak at $\ell=6$, with values three orders of magnitude larger than other values of $\ell$. Extracting the values $\psi_{m6}$ from this snapshot, we constructed the exact solution in Fig~1(a) in the Main Text by letting it have only these non-vanishing expansion coefficients. This results in $\psi_{06} = -4.4$, $\psi_{16} = -2.4 -1.8i$, $\psi_{26} = 0.4 + 1.3i$, $\psi_{36} = 143.9 + 40.8i$, $\psi_{46} = -15.2 + 2.5i$, $\psi_{56} = -2.9 - 20.0i$, $\psi_{66} = 40.8 + 14.5i$, and $\psi_{-m\ell} = (-1)^m \overline{\psi}_{m\ell}$. 
The exact solution in Fig~1(b) in the Main Text has $\ell=30$ and non-vanishing expansion coefficients $\psi_{0\ell} = 8\sqrt{95993978542907}$, $\psi_{\pm 5\ell} = \pm 6\sqrt{2266150070307981}$, $\psi_{10\ell} = 369\sqrt{6048837670715}$, $\psi_{\pm 15\ell} = \pm 496\sqrt{5224419474285}$, $\psi_{20\ell} = 6483\sqrt{5330890838}$, $\psi_{\pm 25\ell} = \pm 30502\sqrt{8224777}$, $\psi_{30\ell} = 79290599\sqrt{77}$.

\section{Derivation of characteristic parameters} 
Before non-dimensionalizing the time scale, the linearized equation for the stream function on a sphere of radius $R$ has the form
\begin{equation}
\partial_t \omega = f(\Delta + 4K)(\Delta + 2K)\omega.
\end{equation}
Writing the shorthand $\delta = (\ell(\ell+1)-4)/R^2$ and moving to the spherical harmonics basis, this becomes 
\begin{align}
\partial_t \omega_{m\ell} &= -\left( \delta + \frac{2}{R^2} \right)( \Gamma_0 + \Gamma_2\delta + \Gamma_4\delta^2)\omega_{m\ell}.
\end{align}
The time-dependent solution of this is
\begin{equation}\label{eq:timeHarmonicSoln}
\omega_{m\ell}(t) = \omega_{m\ell}(0)e^{\sigma_{m\ell}t},
\end{equation}
where
\begin{equation}
\sigma_{m\ell} = -\left( \delta + \frac{2}{R^2} \right)( \Gamma_0 + \Gamma_2\delta + \Gamma_4\delta^2).
\end{equation}
The flow of Eq.~\eqref{eq:timeHarmonicSoln} exhibits a characteristic wave number $\ell_c$ given by the maximum of $\sigma_{m\ell}$. We approximate this maximum by the maxmimum of the function
\begin{equation}
-( \Gamma_0 + \Gamma_2\delta + \Gamma_4\delta^2).
\end{equation}
This results in $\delta_c = -\frac{\Gamma_2}{2\Gamma_4}$ and
\begin{equation}
\ell_c = \frac{1}{2}\left( -1 + 2\sqrt{\frac{17}{4} - \frac{\Gamma_2}{2\Gamma_4}R^2}\right).
\end{equation}
There is then an associated wavelength $\lambda_c = 2\pi R/\ell_c$, corresponding to two vortices - each of characteristic diameter
\begin{equation}
\Lambda = \frac{2\pi R}{2\sqrt{\frac{17}{4} - \frac{\Gamma_2}{2\Gamma_4}R^2} - 1}.
\end{equation}
Next, at the characteristic wave number $\ell_c$, the flow of Eq.~\eqref{eq:timeHarmonicSoln} has the characteristic time-scale
\begin{equation}
\tau = \sigma_{m\ell_c}^{-1} = \left[ \left(\frac{\Gamma_2}{2\Gamma_4} - \frac{2}{R^2} \right)\cdot \left(\Gamma_0 - \frac{\Gamma_2^2}{4\Gamma_4}\right)\right]^{-1},
\end{equation}
and a characteristic spectral bandwidth $\kappa$, defined by
\be
\kappa=\frac{\ell_+-\ell_-}{R},
\ee
where $\ell_\pm$ are the $\ell$-values corresponding to the positive roots $\delta_{\pm}$ of
\be
\Gamma_0 + \Gamma_2\delta + \Gamma_4\delta^2 = 0.
\ee
We have (remember that $-\Gamma_2>0$)
\be
\delta_\pm=\f{1}{2\Gamma_4}(-\Gamma_2\pm \sqrt{\Gamma_2^2-4\Gamma_0\Gamma_4}).
\ee

This results in
\be
\ell_\pm=\f{1}{2}(-1+ \sqrt{17+4\delta_\pm R^2}),
\ee
so the bandwidth is
\be
\kappa=\f{1}{2R}\Big\{\sqrt{17+4\delta_+R^2}-\sqrt{17+4\delta_-R^2}\Big\}.
\ee
We further manipulate
\be
\begin{aligned}
\kappa =&
\Big\{ \frac{17}{2R^2}+\delta_+ +\delta_- \\
&-\frac{1}{2R^2}\Big[ (17+4\delta_+R^2)(17+4\delta_-R^2) \Big] ^{\frac{1}{2}}\Big\}^{\frac{1}{2}} \\
=&
\Big\{ \frac{17}{2R^2}+\delta_+ +\delta_-\\
& - \left[ \frac{17^2}{4R^4}+\frac{17}{R^2}(\delta_+ +\delta_-)+4\delta_+\delta_-  \right]^{\frac{1}{2}}\Big\}^{\frac{1}{2}}  \\
=&
\sqrt{\frac{17}{2R^2}-\f{\Gamma_2}{\Gamma_4}-2\sqrt{\frac{17^2}{16R^4}-\frac{17}{4R^2}\f{\Gamma_2}{\Gamma_4}+\f{\Gamma_0}{\Gamma_4}}}.
\end{aligned} 
\ee

In the limit $R\to\infty$, we recover $\kappa\to \sqrt{-{\Gamma_2}/{\Gamma_4}-2\sqrt{{\Gamma_0}/{\Gamma_4}}}$ which is the expression for the bandwidth in the flat case \cite{slomka2017geometry,slomka2017spontaneous}.

\section{Vortex detection scheme}
We fix a threshold $\alpha_{\omega} \in [0,1]$ and define
\begin{equation}
\widetilde{\omega}(x,t) \!= \! \begin{cases}
0, \text{if}  \min\limits_{x\in S^2} \! \omega(x,t) \! < \!\frac{\omega(x,t)}{\alpha_{\omega}} \!<\! \max\limits_{x\in S^2} \omega(x,t), \\
\omega(x,t), \text{ otherwise}.
\end{cases}
\end{equation}
The number of vortices present on the sphere at time~$t$, $N_{\omega}(\kappa, R ; t)$, is then defined to be the number of connected components of the region $\{ x : \widetilde{\omega}(x,t) \neq 0\}$. Fig.~\ref{fig:SM_detection} demonstrates this procedure. The large-scale branched structure of the vorticity field is captured well after thresholding, justifying this simple vortex detection scheme. 
\par
Next, we characterize the geometrical difference in the behavior of the surface tension chains. Calculating the average surface tension $\overline{\sigma}(t)$ on the sphere, we define a thresholded surface tension by
\begin{equation}
\widetilde{\sigma}(x,t) = \begin{cases}
\overline{\sigma}(t), \text{if} \min\limits_{x\in S^2} \sigma(x,t) \!< \overline{\sigma}(t), \\
\sigma(x,t), \text{ otherwise}.
\end{cases}
\end{equation}
For each connected component of the region where $\widetilde{\sigma}(x,t) > \overline{\sigma}(t)$, we measure its area $A$, together with the area of its boundary pixels ${\partial A}$. The ratio ${\partial A}/A$ is then a measure of the chain-like structure in the tension fields, with a large value signaling a highly branched structure, whereas smaller values indicate less branching.

We denote the Betti number of vortices for a parameter pair $(\kappa, R)$ at time $t$ as $N_{\omega}(\kappa, R ; t)$, and the sum of the ratios ${\partial A}/A$ for every connected component in the region where $\widetilde{\sigma}(x,t) \neq 0$ by $A_{\sigma}(\kappa, R ; t)$. To normalize these quantities, we define a reference value $\kappa_* = 0.3/\Lambda$, corresponding to a flow pattern exhibiting the anomalous turbulent phase, for all measured values of $\Lambda$. With this, we can define a normalized Betti number of vortices as
\be
\label{sme:betti}
\mathrm{Betti}_\omega = \frac{\langle N_{\omega}(\kappa, R ; t) - N_{\omega}(\kappa_*, R ; t) \rangle}{\langle N_{\omega}(\kappa_*, R ; t) \rangle}
\ee
and a relative branching index for the high-tension  areas
\begin{equation}
\mathrm{Branch}_{\sigma} = \frac{\langle A_{\sigma}(\kappa, R ; t) - A_{\sigma}(\kappa_*, R ; t) \rangle}{\langle A_{\sigma}(\kappa_*, R ; t) \rangle},
\end{equation}
where the averages are taken over time after the initial relaxation period.

\begin{figure}
\includegraphics[width=1\textwidth]{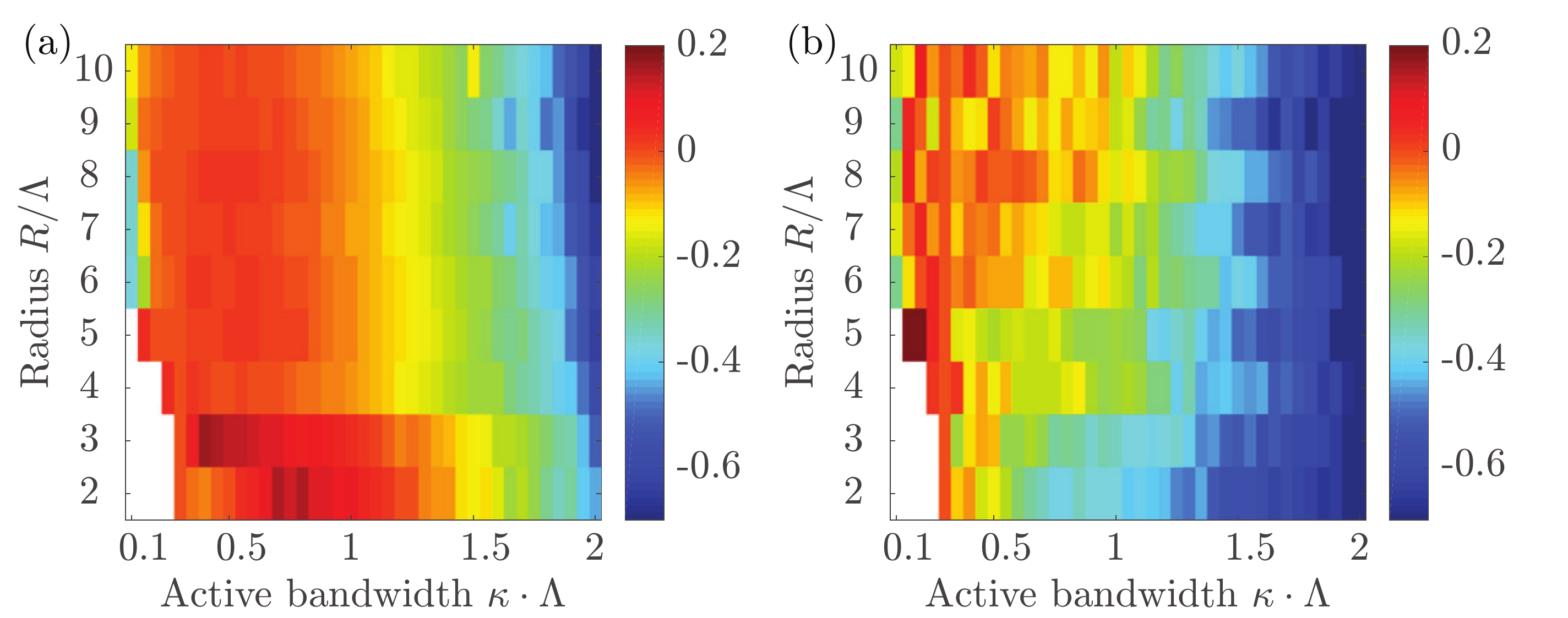}
\caption{Phase diagram for $\alpha_{\omega} = 0.25$ (a) and $\alpha_{\omega} = 0.75$~(b), showing  that qualitative changes in the different turbulent phases are robust with regard to variations in $\alpha_{\omega}$; cf. Fig.~2(a) in the Main Text. Color scales show normalized Betti number defined in Eq.~\eqref{sme:betti}.
\label{fig:SM_phase}} 
\end{figure}
\subsection{Robustness of phase transition to thresholding} 
Fig.~\ref{fig:SM_phase} shows phase diagrams using the thresholding parameters $\alpha_{\omega} = 0.25$ (a) and $\alpha_{\omega} = 0.75$ (b). The phase transition exhibits the same qualitative behavior for these parameter value as compared to Fig.~2(a) in the Main Text, indicating robustness to the method used.

\section{Enstrophy evolution} 
\begin{figure}[h!]
\includegraphics[width=0.85\textwidth]{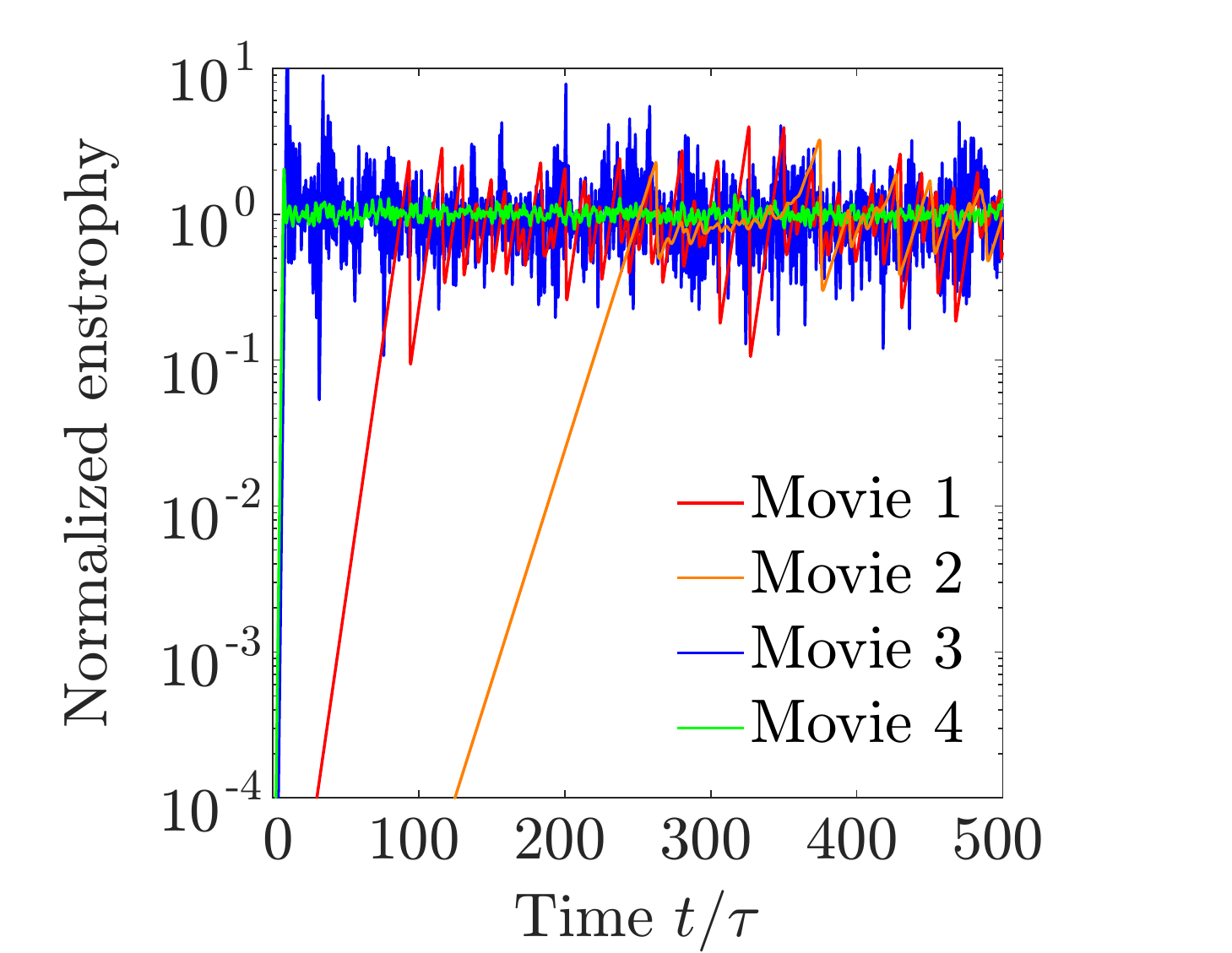}
\caption{Enstrophy normalized by mean after relaxation for each of Movies~1-4.
\label{fig:SM_enstrophy}} 
\end{figure}

Fig.~\ref{fig:SM_enstrophy} shows the evolution of the total enstrophy
\begin{equation}
\int_S \omega^2 d\Omega,
\end{equation}
for each of Movies~1-4. Movies~1 and 2 exhibit a significantly longer relaxation time and their dynamics are given by periods of energy storage interchanged with burst-like movement and corresponding decrease in enstrophy. 

\begin{thebibliography}{69}%
\makeatletter
\providecommand \@ifxundefined [1]{%
 \@ifx{#1\undefined}
}%
\providecommand \@ifnum [1]{%
 \ifnum #1\expandafter \@firstoftwo
 \else \expandafter \@secondoftwo
 \fi
}%
\providecommand \@ifx [1]{%
 \ifx #1\expandafter \@firstoftwo
 \else \expandafter \@secondoftwo
 \fi
}%
\providecommand \natexlab [1]{#1}%
\providecommand \enquote  [1]{``#1''}%
\providecommand \bibnamefont  [1]{#1}%
\providecommand \bibfnamefont [1]{#1}%
\providecommand \citenamefont [1]{#1}%
\providecommand \href@noop [0]{\@secondoftwo}%
\providecommand \href [0]{\begingroup \@sanitize@url \@href}%
\providecommand \@href[1]{\@@startlink{#1}\@@href}%
\providecommand \@@href[1]{\endgroup#1\@@endlink}%
\providecommand \@sanitize@url [0]{\catcode `\\12\catcode `\$12\catcode
  `\&12\catcode `\#12\catcode `\^12\catcode `\_12\catcode `\%12\relax}%
\providecommand \@@startlink[1]{}%
\providecommand \@@endlink[0]{}%
\providecommand \url  [0]{\begingroup\@sanitize@url \@url }%
\providecommand \@url [1]{\endgroup\@href {#1}{\urlprefix }}%
\providecommand \urlprefix  [0]{URL }%
\providecommand \Eprint [0]{\href }%
\providecommand \doibase [0]{http://dx.doi.org/}%
\providecommand \selectlanguage [0]{\@gobble}%
\providecommand \bibinfo  [0]{\@secondoftwo}%
\providecommand \bibfield  [0]{\@secondoftwo}%
\providecommand \translation [1]{[#1]}%
\providecommand \BibitemOpen [0]{}%
\providecommand \bibitemStop [0]{}%
\providecommand \bibitemNoStop [0]{.\EOS\space}%
\providecommand \EOS [0]{\spacefactor3000\relax}%
\providecommand \BibitemShut  [1]{\csname bibitem#1\endcsname}%
\let\auto@bib@innerbib\@empty
\bibitem [{\citenamefont {Cranmer}\ and\ \citenamefont {van
  Ballegooijen}(2005)}]{2005Cranmer}%
  \BibitemOpen
  \bibfield  {author} {\bibinfo {author} {\bibfnamefont {S.}~\bibnamefont
  {Cranmer}}\ and\ \bibinfo {author} {\bibfnamefont {A.}~\bibnamefont {van
  Ballegooijen}},\ }\href@noop {} {\bibfield  {journal} {\bibinfo  {journal}
  {Astrophys. J. Suppl. Ser.}\ }\textbf {\bibinfo {volume} {156}},\ \bibinfo
  {pages} {265} (\bibinfo {year} {2005})}\BibitemShut {NoStop}%
\bibitem [{\citenamefont {Keber}\ \emph {et~al.}(2014)\citenamefont {Keber},
  \citenamefont {Loiseau}, \citenamefont {Sanchez}, \citenamefont {DeCamp},
  \citenamefont {Giomi}, \citenamefont {Bowick}, \citenamefont {Marchetti},
  \citenamefont {Dogic},\ and\ \citenamefont {Bausch}}]{keber2014topology}%
  \BibitemOpen
  \bibfield  {author} {\bibinfo {author} {\bibfnamefont {F.~C.}\ \bibnamefont
  {Keber}}, \bibinfo {author} {\bibfnamefont {E.}~\bibnamefont {Loiseau}},
  \bibinfo {author} {\bibfnamefont {T.}~\bibnamefont {Sanchez}}, \bibinfo
  {author} {\bibfnamefont {S.~J.}\ \bibnamefont {DeCamp}}, \bibinfo {author}
  {\bibfnamefont {L.}~\bibnamefont {Giomi}}, \bibinfo {author} {\bibfnamefont
  {M.~J.}\ \bibnamefont {Bowick}}, \bibinfo {author} {\bibfnamefont {M.~C.}\
  \bibnamefont {Marchetti}}, \bibinfo {author} {\bibfnamefont {Z.}~\bibnamefont
  {Dogic}}, \ and\ \bibinfo {author} {\bibfnamefont {A.~R.}\ \bibnamefont
  {Bausch}},\ }\href@noop {} {\bibfield  {journal} {\bibinfo  {journal}
  {Science}\ }\textbf {\bibinfo {volume} {345}},\ \bibinfo {pages} {1135}
  (\bibinfo {year} {2014})}\BibitemShut {NoStop}%
\bibitem [{\citenamefont {Zhang}\ \emph {et~al.}(2016)\citenamefont {Zhang},
  \citenamefont {Zhou}, \citenamefont {Rahimi},\ and\ \citenamefont
  {de~Pablo}}]{Zhang:2016aa}%
  \BibitemOpen
  \bibfield  {author} {\bibinfo {author} {\bibfnamefont {R.}~\bibnamefont
  {Zhang}}, \bibinfo {author} {\bibfnamefont {Y.}~\bibnamefont {Zhou}},
  \bibinfo {author} {\bibfnamefont {M.}~\bibnamefont {Rahimi}}, \ and\ \bibinfo
  {author} {\bibfnamefont {J.~J.}\ \bibnamefont {de~Pablo}},\ }\href@noop {}
  {\bibfield  {journal} {\bibinfo  {journal} {Nat. Commun.}\ }\textbf {\bibinfo
  {volume} {7}},\ \bibinfo {pages} {13483} (\bibinfo {year}
  {2016})}\BibitemShut {NoStop}%
\bibitem [{\citenamefont {Sipos}\ \emph {et~al.}(2015)\citenamefont {Sipos},
  \citenamefont {Nagy}, \citenamefont {Di~Leonardo},\ and\ \citenamefont
  {Galajda}}]{sipos2015hydrodynamic}%
  \BibitemOpen
  \bibfield  {author} {\bibinfo {author} {\bibfnamefont {O.}~\bibnamefont
  {Sipos}}, \bibinfo {author} {\bibfnamefont {K.}~\bibnamefont {Nagy}},
  \bibinfo {author} {\bibfnamefont {R.}~\bibnamefont {Di~Leonardo}}, \ and\
  \bibinfo {author} {\bibfnamefont {P.}~\bibnamefont {Galajda}},\ }\href@noop
  {} {\bibfield  {journal} {\bibinfo  {journal} {Phys. Rev. Lett.}\ }\textbf
  {\bibinfo {volume} {114}},\ \bibinfo {pages} {258104} (\bibinfo {year}
  {2015})}\BibitemShut {NoStop}%
\bibitem [{\citenamefont {Chang}\ \emph {et~al.}(2015)\citenamefont {Chang},
  \citenamefont {Fragkopoulos}, \citenamefont {Marquez}, \citenamefont {Kim},
  \citenamefont {Angelini},\ and\ \citenamefont
  {Fernandez-Nieves}}]{2015Chang_NJP}%
  \BibitemOpen
  \bibfield  {author} {\bibinfo {author} {\bibfnamefont {Y.-W.}\ \bibnamefont
  {Chang}}, \bibinfo {author} {\bibfnamefont {A.~A.}\ \bibnamefont
  {Fragkopoulos}}, \bibinfo {author} {\bibfnamefont {S.~M.}\ \bibnamefont
  {Marquez}}, \bibinfo {author} {\bibfnamefont {H.~D.}\ \bibnamefont {Kim}},
  \bibinfo {author} {\bibfnamefont {T.~E.}\ \bibnamefont {Angelini}}, \ and\
  \bibinfo {author} {\bibfnamefont {A.}~\bibnamefont {Fernandez-Nieves}},\
  }\href@noop {} {\bibfield  {journal} {\bibinfo  {journal} {New J. Phys.}\
  }\textbf {\bibinfo {volume} {17}},\ \bibinfo {pages} {033017} (\bibinfo
  {year} {2015})}\BibitemShut {NoStop}%
\bibitem [{\citenamefont {Das}\ and\ \citenamefont
  {Mukherjee}(2007)}]{Das2007}%
  \BibitemOpen
  \bibfield  {author} {\bibinfo {author} {\bibfnamefont {K.}~\bibnamefont
  {Das}}\ and\ \bibinfo {author} {\bibfnamefont {A.~K.}\ \bibnamefont
  {Mukherjee}},\ }\href@noop {} {\bibfield  {journal} {\bibinfo  {journal}
  {Bioresour. Technol.}\ }\textbf {\bibinfo {volume} {98}},\ \bibinfo {pages}
  {1339} (\bibinfo {year} {2007})}\BibitemShut {NoStop}%
\bibitem [{\citenamefont {Bold}(1949)}]{Bold:1949aa}%
  \BibitemOpen
  \bibfield  {author} {\bibinfo {author} {\bibfnamefont {H.~C.}\ \bibnamefont
  {Bold}},\ }\href {http://www.jstor.org/stable/2482218} {\bibfield  {journal}
  {\bibinfo  {journal} {Bulletin of the Torrey Botanical Club}\ }\textbf
  {\bibinfo {volume} {76}},\ \bibinfo {pages} {101} (\bibinfo {year}
  {1949})}\BibitemShut {NoStop}%
\bibitem [{\citenamefont {Costerton}\ \emph {et~al.}(1999)\citenamefont
  {Costerton}, \citenamefont {Stewart},\ and\ \citenamefont
  {Greenberg}}]{Costerton:1999aa}%
  \BibitemOpen
  \bibfield  {author} {\bibinfo {author} {\bibfnamefont {J.~W.}\ \bibnamefont
  {Costerton}}, \bibinfo {author} {\bibfnamefont {P.~S.}\ \bibnamefont
  {Stewart}}, \ and\ \bibinfo {author} {\bibfnamefont {E.~P.}\ \bibnamefont
  {Greenberg}},\ }\href
  {http://www.sciencemag.org/content/284/5418/1318.abstract N2 - Bacteria that
  attach to surfaces aggregate in a hydrated polymeric matrix of their own
  synthesis to form biofilms. Formation of these sessile communities and their
  inherent resistance to antimicrobial agents are at the root of many
  persistent and chronic bacterial infections. Studies of biofilms have
  revealed differentiated, structured groups of cells with community
  properties. Recent advances in our understanding of the genetic and molecular
  basis of bacterial community behavior point to therapeutic targets that may
  provide a means for the control of biofilm infections.} {\bibfield  {journal}
  {\bibinfo  {journal} {Science}\ }\textbf {\bibinfo {volume} {284}},\ \bibinfo
  {pages} {1318} (\bibinfo {year} {1999})}\BibitemShut {NoStop}%
\bibitem [{\citenamefont {Rosenberg}\ \emph {et~al.}(1992)\citenamefont
  {Rosenberg}, \citenamefont {Legmann}, \citenamefont {Kushmaro}, \citenamefont
  {Taube}, \citenamefont {Adler},\ and\ \citenamefont {Ron}}]{1992Rosenberg}%
  \BibitemOpen
  \bibfield  {author} {\bibinfo {author} {\bibfnamefont {E.}~\bibnamefont
  {Rosenberg}}, \bibinfo {author} {\bibfnamefont {R.}~\bibnamefont {Legmann}},
  \bibinfo {author} {\bibfnamefont {A.}~\bibnamefont {Kushmaro}}, \bibinfo
  {author} {\bibfnamefont {R.}~\bibnamefont {Taube}}, \bibinfo {author}
  {\bibfnamefont {E.}~\bibnamefont {Adler}}, \ and\ \bibinfo {author}
  {\bibfnamefont {E.~Z.}\ \bibnamefont {Ron}},\ }\href@noop {} {\bibfield
  {journal} {\bibinfo  {journal} {Biodegradation}\ }\textbf {\bibinfo {volume}
  {3}},\ \bibinfo {pages} {337} (\bibinfo {year} {1992})}\BibitemShut {NoStop}%
\bibitem [{\citenamefont {Rosenberg}(2006)}]{2006Rosenberg}%
  \BibitemOpen
  \bibfield  {author} {\bibinfo {author} {\bibfnamefont {M.}~\bibnamefont
  {Rosenberg}},\ }\href@noop {} {\bibfield  {journal} {\bibinfo  {journal}
  {FEMS Microbiol. Lett.}\ }\textbf {\bibinfo {volume} {262}},\ \bibinfo
  {pages} {129} (\bibinfo {year} {2006})}\BibitemShut {NoStop}%
\bibitem [{\citenamefont {Rosenberg}\ \emph {et~al.}(2013)\citenamefont
  {Rosenberg}, \citenamefont {DeLong}, \citenamefont {Lory}, \citenamefont
  {Stackebrandt},\ and\ \citenamefont {Thompson}}]{2013Rosenberg}%
  \BibitemOpen
  \bibfield  {author} {\bibinfo {author} {\bibfnamefont {E.}~\bibnamefont
  {Rosenberg}}, \bibinfo {author} {\bibfnamefont {E.~F.}\ \bibnamefont
  {DeLong}}, \bibinfo {author} {\bibfnamefont {S.}~\bibnamefont {Lory}},
  \bibinfo {author} {\bibfnamefont {E.}~\bibnamefont {Stackebrandt}}, \ and\
  \bibinfo {author} {\bibfnamefont {F.}~\bibnamefont {Thompson}},\ }\enquote
  {\bibinfo {title} {Hydrocarbon-oxidizing bacteria},}\ in\ \href@noop {}
  {\emph {\bibinfo {booktitle} {The Prokaryotes}}}\ (\bibinfo  {publisher}
  {Springer Berlin Heidelberg},\ \bibinfo {year} {2013})\ pp.\ \bibinfo {pages}
  {201--214}\BibitemShut {NoStop}%
\bibitem [{\citenamefont {Vicsek}\ \emph {et~al.}(1995)\citenamefont {Vicsek},
  \citenamefont {Czir{\'o}k}, \citenamefont {Ben-Jacob}, \citenamefont
  {Cohen},\ and\ \citenamefont {Shochet}}]{vicsek1995novel}%
  \BibitemOpen
  \bibfield  {author} {\bibinfo {author} {\bibfnamefont {T.}~\bibnamefont
  {Vicsek}}, \bibinfo {author} {\bibfnamefont {A.}~\bibnamefont {Czir{\'o}k}},
  \bibinfo {author} {\bibfnamefont {E.}~\bibnamefont {Ben-Jacob}}, \bibinfo
  {author} {\bibfnamefont {I.}~\bibnamefont {Cohen}}, \ and\ \bibinfo {author}
  {\bibfnamefont {O.}~\bibnamefont {Shochet}},\ }\href@noop {} {\bibfield
  {journal} {\bibinfo  {journal} {Phys. Rev. Lett.}\ }\textbf {\bibinfo
  {volume} {75}},\ \bibinfo {pages} {1226} (\bibinfo {year}
  {1995})}\BibitemShut {NoStop}%
\bibitem [{\citenamefont {Toner}\ and\ \citenamefont
  {Tu}(1995)}]{toner1995long}%
  \BibitemOpen
  \bibfield  {author} {\bibinfo {author} {\bibfnamefont {J.}~\bibnamefont
  {Toner}}\ and\ \bibinfo {author} {\bibfnamefont {Y.}~\bibnamefont {Tu}},\
  }\href@noop {} {\bibfield  {journal} {\bibinfo  {journal} {Phys. Rev. Lett.}\
  }\textbf {\bibinfo {volume} {75}},\ \bibinfo {pages} {4326} (\bibinfo {year}
  {1995})}\BibitemShut {NoStop}%
\bibitem [{\citenamefont {Baskaran}\ and\ \citenamefont
  {Marchetti}(2009)}]{baskaran2009statistical}%
  \BibitemOpen
  \bibfield  {author} {\bibinfo {author} {\bibfnamefont {A.}~\bibnamefont
  {Baskaran}}\ and\ \bibinfo {author} {\bibfnamefont {M.~C.}\ \bibnamefont
  {Marchetti}},\ }\href@noop {} {\bibfield  {journal} {\bibinfo  {journal}
  {Proc. Nat. Acad. Sci. U.S.A.}\ }\textbf {\bibinfo {volume} {106}},\ \bibinfo
  {pages} {15567} (\bibinfo {year} {2009})}\BibitemShut {NoStop}%
\bibitem [{\citenamefont {Koch}\ and\ \citenamefont
  {Subramanian}(2011)}]{koch2011collective}%
  \BibitemOpen
  \bibfield  {author} {\bibinfo {author} {\bibfnamefont {D.~L.}\ \bibnamefont
  {Koch}}\ and\ \bibinfo {author} {\bibfnamefont {G.}~\bibnamefont
  {Subramanian}},\ }\href@noop {} {\bibfield  {journal} {\bibinfo  {journal}
  {Annu. Rev. Fluid Mech.}\ }\textbf {\bibinfo {volume} {43}},\ \bibinfo
  {pages} {637} (\bibinfo {year} {2011})}\BibitemShut {NoStop}%
\bibitem [{\citenamefont {Ramaswamy}(2010)}]{ramaswamy2010mechanics}%
  \BibitemOpen
  \bibfield  {author} {\bibinfo {author} {\bibfnamefont {S.}~\bibnamefont
  {Ramaswamy}},\ }\href@noop {} {\bibfield  {journal} {\bibinfo  {journal}
  {Annu. Rev. Condens. Matter Phys.}\ }\textbf {\bibinfo {volume} {1}},\
  \bibinfo {pages} {323} (\bibinfo {year} {2010})}\BibitemShut {NoStop}%
\bibitem [{\citenamefont {Marchetti}\ \emph {et~al.}(2013)\citenamefont
  {Marchetti}, \citenamefont {Joanny}, \citenamefont {Ramaswamy}, \citenamefont
  {Liverpool}, \citenamefont {Prost}, \citenamefont {Rao},\ and\ \citenamefont
  {Simha}}]{marchetti2013hydrodynamics}%
  \BibitemOpen
  \bibfield  {author} {\bibinfo {author} {\bibfnamefont {M.~C.}\ \bibnamefont
  {Marchetti}}, \bibinfo {author} {\bibfnamefont {J.}~\bibnamefont {Joanny}},
  \bibinfo {author} {\bibfnamefont {S.}~\bibnamefont {Ramaswamy}}, \bibinfo
  {author} {\bibfnamefont {T.}~\bibnamefont {Liverpool}}, \bibinfo {author}
  {\bibfnamefont {J.}~\bibnamefont {Prost}}, \bibinfo {author} {\bibfnamefont
  {M.}~\bibnamefont {Rao}}, \ and\ \bibinfo {author} {\bibfnamefont {R.~A.}\
  \bibnamefont {Simha}},\ }\href@noop {} {\bibfield  {journal} {\bibinfo
  {journal} {Rev. Mod. Phys.}\ }\textbf {\bibinfo {volume} {85}},\ \bibinfo
  {pages} {1143} (\bibinfo {year} {2013})}\BibitemShut {NoStop}%
\bibitem [{\citenamefont {Kruse}\ \emph {et~al.}(2004)\citenamefont {Kruse},
  \citenamefont {Joanny}, \citenamefont {J{\"u}licher}, \citenamefont {Prost},\
  and\ \citenamefont {Sekimoto}}]{kruse2004asters}%
  \BibitemOpen
  \bibfield  {author} {\bibinfo {author} {\bibfnamefont {K.}~\bibnamefont
  {Kruse}}, \bibinfo {author} {\bibfnamefont {J.-F.}\ \bibnamefont {Joanny}},
  \bibinfo {author} {\bibfnamefont {F.}~\bibnamefont {J{\"u}licher}}, \bibinfo
  {author} {\bibfnamefont {J.}~\bibnamefont {Prost}}, \ and\ \bibinfo {author}
  {\bibfnamefont {K.}~\bibnamefont {Sekimoto}},\ }\href@noop {} {\bibfield
  {journal} {\bibinfo  {journal} {Phys. Rev. Lett.}\ }\textbf {\bibinfo
  {volume} {92}},\ \bibinfo {pages} {078101} (\bibinfo {year}
  {2004})}\BibitemShut {NoStop}%
\bibitem [{\citenamefont {Wolgemuth}(2008)}]{wolgemuth2008collective}%
  \BibitemOpen
  \bibfield  {author} {\bibinfo {author} {\bibfnamefont {C.~W.}\ \bibnamefont
  {Wolgemuth}},\ }\href@noop {} {\bibfield  {journal} {\bibinfo  {journal}
  {Biophys. J.}\ }\textbf {\bibinfo {volume} {95}},\ \bibinfo {pages} {1564}
  (\bibinfo {year} {2008})}\BibitemShut {NoStop}%
\bibitem [{\citenamefont {Saintillan}\ and\ \citenamefont
  {Shelley}(2008)}]{saintillan2008instabilities}%
  \BibitemOpen
  \bibfield  {author} {\bibinfo {author} {\bibfnamefont {D.}~\bibnamefont
  {Saintillan}}\ and\ \bibinfo {author} {\bibfnamefont {M.~J.}\ \bibnamefont
  {Shelley}},\ }\href@noop {} {\bibfield  {journal} {\bibinfo  {journal} {Phys.
  Fluids}\ }\textbf {\bibinfo {volume} {20}},\ \bibinfo {pages} {123304}
  (\bibinfo {year} {2008})}\BibitemShut {NoStop}%
\bibitem [{\citenamefont {Peshkov}\ \emph {et~al.}(2012)\citenamefont
  {Peshkov}, \citenamefont {Aranson}, \citenamefont {Bertin}, \citenamefont
  {Chat{\'e}},\ and\ \citenamefont {Ginelli}}]{peshkov2012nonlinear}%
  \BibitemOpen
  \bibfield  {author} {\bibinfo {author} {\bibfnamefont {A.}~\bibnamefont
  {Peshkov}}, \bibinfo {author} {\bibfnamefont {I.~S.}\ \bibnamefont
  {Aranson}}, \bibinfo {author} {\bibfnamefont {E.}~\bibnamefont {Bertin}},
  \bibinfo {author} {\bibfnamefont {H.}~\bibnamefont {Chat{\'e}}}, \ and\
  \bibinfo {author} {\bibfnamefont {F.}~\bibnamefont {Ginelli}},\ }\href@noop
  {} {\bibfield  {journal} {\bibinfo  {journal} {Phys. Rev. Lett.}\ }\textbf
  {\bibinfo {volume} {109}},\ \bibinfo {pages} {268701} (\bibinfo {year}
  {2012})}\BibitemShut {NoStop}%
\bibitem [{\citenamefont {Brotto}\ \emph {et~al.}(2013)\citenamefont {Brotto},
  \citenamefont {Caussin}, \citenamefont {Lauga},\ and\ \citenamefont
  {Bartolo}}]{brotto2013hydrodynamics}%
  \BibitemOpen
  \bibfield  {author} {\bibinfo {author} {\bibfnamefont {T.}~\bibnamefont
  {Brotto}}, \bibinfo {author} {\bibfnamefont {J.-B.}\ \bibnamefont {Caussin}},
  \bibinfo {author} {\bibfnamefont {E.}~\bibnamefont {Lauga}}, \ and\ \bibinfo
  {author} {\bibfnamefont {D.}~\bibnamefont {Bartolo}},\ }\href@noop {}
  {\bibfield  {journal} {\bibinfo  {journal} {Phys. Rev. Lett.}\ }\textbf
  {\bibinfo {volume} {110}},\ \bibinfo {pages} {038101} (\bibinfo {year}
  {2013})}\BibitemShut {NoStop}%
\bibitem [{\citenamefont {Thampi}\ \emph {et~al.}(2013)\citenamefont {Thampi},
  \citenamefont {Golestanian},\ and\ \citenamefont
  {Yeomans}}]{thampi2013velocity}%
  \BibitemOpen
  \bibfield  {author} {\bibinfo {author} {\bibfnamefont {S.~P.}\ \bibnamefont
  {Thampi}}, \bibinfo {author} {\bibfnamefont {R.}~\bibnamefont {Golestanian}},
  \ and\ \bibinfo {author} {\bibfnamefont {J.~M.}\ \bibnamefont {Yeomans}},\
  }\href@noop {} {\bibfield  {journal} {\bibinfo  {journal} {Phys. Rev. Lett.}\
  }\textbf {\bibinfo {volume} {111}},\ \bibinfo {pages} {118101} (\bibinfo
  {year} {2013})}\BibitemShut {NoStop}%
\bibitem [{\citenamefont {Mendelson}\ \emph {et~al.}(1999)\citenamefont
  {Mendelson}, \citenamefont {Bourque}, \citenamefont {Wilkening},
  \citenamefont {Anderson},\ and\ \citenamefont
  {Watkins}}]{mendelson1999organized}%
  \BibitemOpen
  \bibfield  {author} {\bibinfo {author} {\bibfnamefont {N.~H.}\ \bibnamefont
  {Mendelson}}, \bibinfo {author} {\bibfnamefont {A.}~\bibnamefont {Bourque}},
  \bibinfo {author} {\bibfnamefont {K.}~\bibnamefont {Wilkening}}, \bibinfo
  {author} {\bibfnamefont {K.~R.}\ \bibnamefont {Anderson}}, \ and\ \bibinfo
  {author} {\bibfnamefont {J.~C.}\ \bibnamefont {Watkins}},\ }\href@noop {}
  {\bibfield  {journal} {\bibinfo  {journal} {J. Bacteriol.}\ }\textbf
  {\bibinfo {volume} {181}},\ \bibinfo {pages} {600} (\bibinfo {year}
  {1999})}\BibitemShut {NoStop}%
\bibitem [{\citenamefont {Steager}\ \emph {et~al.}(2008)\citenamefont
  {Steager}, \citenamefont {Kim},\ and\ \citenamefont
  {Kim}}]{steager2008dynamics}%
  \BibitemOpen
  \bibfield  {author} {\bibinfo {author} {\bibfnamefont {E.~B.}\ \bibnamefont
  {Steager}}, \bibinfo {author} {\bibfnamefont {C.-B.}\ \bibnamefont {Kim}}, \
  and\ \bibinfo {author} {\bibfnamefont {M.~J.}\ \bibnamefont {Kim}},\
  }\href@noop {} {\bibfield  {journal} {\bibinfo  {journal} {Phys. Fluids}\
  }\textbf {\bibinfo {volume} {20}},\ \bibinfo {pages} {073601} (\bibinfo
  {year} {2008})}\BibitemShut {NoStop}%
\bibitem [{\citenamefont {Zhang}\ \emph {et~al.}(2010)\citenamefont {Zhang},
  \citenamefont {Turner},\ and\ \citenamefont {Berg}}]{zhang2010upper}%
  \BibitemOpen
  \bibfield  {author} {\bibinfo {author} {\bibfnamefont {R.}~\bibnamefont
  {Zhang}}, \bibinfo {author} {\bibfnamefont {L.}~\bibnamefont {Turner}}, \
  and\ \bibinfo {author} {\bibfnamefont {H.~C.}\ \bibnamefont {Berg}},\
  }\href@noop {} {\bibfield  {journal} {\bibinfo  {journal} {Proc. Nat. Acad.
  Sci. U.S.A.}\ }\textbf {\bibinfo {volume} {107}},\ \bibinfo {pages} {288}
  (\bibinfo {year} {2010})}\BibitemShut {NoStop}%
\bibitem [{\citenamefont {Sokolov}\ \emph {et~al.}(2007)\citenamefont
  {Sokolov}, \citenamefont {Aranson}, \citenamefont {Kessler},\ and\
  \citenamefont {Goldstein}}]{sokolov2007concentration}%
  \BibitemOpen
  \bibfield  {author} {\bibinfo {author} {\bibfnamefont {A.}~\bibnamefont
  {Sokolov}}, \bibinfo {author} {\bibfnamefont {I.~S.}\ \bibnamefont
  {Aranson}}, \bibinfo {author} {\bibfnamefont {J.~O.}\ \bibnamefont
  {Kessler}}, \ and\ \bibinfo {author} {\bibfnamefont {R.~E.}\ \bibnamefont
  {Goldstein}},\ }\href@noop {} {\bibfield  {journal} {\bibinfo  {journal}
  {Phys. Rev. Lett.}\ }\textbf {\bibinfo {volume} {98}},\ \bibinfo {pages}
  {158102} (\bibinfo {year} {2007})}\BibitemShut {NoStop}%
\bibitem [{\citenamefont {Dombrowski}\ \emph {et~al.}(2004)\citenamefont
  {Dombrowski}, \citenamefont {Cisneros}, \citenamefont {Chatkaew},
  \citenamefont {Goldstein},\ and\ \citenamefont
  {Kessler}}]{dombrowski2004self}%
  \BibitemOpen
  \bibfield  {author} {\bibinfo {author} {\bibfnamefont {C.}~\bibnamefont
  {Dombrowski}}, \bibinfo {author} {\bibfnamefont {L.}~\bibnamefont
  {Cisneros}}, \bibinfo {author} {\bibfnamefont {S.}~\bibnamefont {Chatkaew}},
  \bibinfo {author} {\bibfnamefont {R.~E.}\ \bibnamefont {Goldstein}}, \ and\
  \bibinfo {author} {\bibfnamefont {J.~O.}\ \bibnamefont {Kessler}},\
  }\href@noop {} {\bibfield  {journal} {\bibinfo  {journal} {Phys. Rev. Lett.}\
  }\textbf {\bibinfo {volume} {93}},\ \bibinfo {pages} {098103} (\bibinfo
  {year} {2004})}\BibitemShut {NoStop}%
\bibitem [{\citenamefont {Wu}\ \emph {et~al.}(2006)\citenamefont {Wu},
  \citenamefont {Roberts}, \citenamefont {Kim}, \citenamefont {Koch},\ and\
  \citenamefont {DeLisa}}]{wu2006collective}%
  \BibitemOpen
  \bibfield  {author} {\bibinfo {author} {\bibfnamefont {M.}~\bibnamefont
  {Wu}}, \bibinfo {author} {\bibfnamefont {J.~W.}\ \bibnamefont {Roberts}},
  \bibinfo {author} {\bibfnamefont {S.}~\bibnamefont {Kim}}, \bibinfo {author}
  {\bibfnamefont {D.~L.}\ \bibnamefont {Koch}}, \ and\ \bibinfo {author}
  {\bibfnamefont {M.~P.}\ \bibnamefont {DeLisa}},\ }\href@noop {} {\bibfield
  {journal} {\bibinfo  {journal} {Appl. Environ. Microbiol.}\ }\textbf
  {\bibinfo {volume} {72}},\ \bibinfo {pages} {4987} (\bibinfo {year}
  {2006})}\BibitemShut {NoStop}%
\bibitem [{\citenamefont {Dunkel}\ \emph {et~al.}(2013)\citenamefont {Dunkel},
  \citenamefont {Heidenreich}, \citenamefont {Drescher}, \citenamefont
  {Wensink}, \citenamefont {B{\"a}r},\ and\ \citenamefont
  {Goldstein}}]{dunkel2013fluid}%
  \BibitemOpen
  \bibfield  {author} {\bibinfo {author} {\bibfnamefont {J.}~\bibnamefont
  {Dunkel}}, \bibinfo {author} {\bibfnamefont {S.}~\bibnamefont {Heidenreich}},
  \bibinfo {author} {\bibfnamefont {K.}~\bibnamefont {Drescher}}, \bibinfo
  {author} {\bibfnamefont {H.~H.}\ \bibnamefont {Wensink}}, \bibinfo {author}
  {\bibfnamefont {M.}~\bibnamefont {B{\"a}r}}, \ and\ \bibinfo {author}
  {\bibfnamefont {R.~E.}\ \bibnamefont {Goldstein}},\ }\href@noop {} {\bibfield
   {journal} {\bibinfo  {journal} {Phys. Rev. Lett.}\ }\textbf {\bibinfo
  {volume} {110}},\ \bibinfo {pages} {228102} (\bibinfo {year}
  {2013})}\BibitemShut {NoStop}%
\bibitem [{\citenamefont {Sknepnek}\ and\ \citenamefont
  {Henkes}(2015)}]{sknepnek2015active}%
  \BibitemOpen
  \bibfield  {author} {\bibinfo {author} {\bibfnamefont {R.}~\bibnamefont
  {Sknepnek}}\ and\ \bibinfo {author} {\bibfnamefont {S.}~\bibnamefont
  {Henkes}},\ }\href@noop {} {\bibfield  {journal} {\bibinfo  {journal} {Phys.
  Rev. E}\ }\textbf {\bibinfo {volume} {91}},\ \bibinfo {pages} {022306}
  (\bibinfo {year} {2015})}\BibitemShut {NoStop}%
\bibitem [{\citenamefont {Fily}\ \emph {et~al.}(2016)\citenamefont {Fily},
  \citenamefont {Baskaran},\ and\ \citenamefont {Hagan}}]{fily2016active}%
  \BibitemOpen
  \bibfield  {author} {\bibinfo {author} {\bibfnamefont {Y.}~\bibnamefont
  {Fily}}, \bibinfo {author} {\bibfnamefont {A.}~\bibnamefont {Baskaran}}, \
  and\ \bibinfo {author} {\bibfnamefont {M.}~\bibnamefont {Hagan}},\
  }\href@noop {} {\bibfield  {journal} {\bibinfo  {journal} {arXiv:1601.00324}\
  } (\bibinfo {year} {2016})}\BibitemShut {NoStop}%
\bibitem [{\citenamefont {Janssen}\ \emph {et~al.}(2017)\citenamefont
  {Janssen}, \citenamefont {Kaiser},\ and\ \citenamefont
  {L{\"o}wen}}]{janssen2017aging}%
  \BibitemOpen
  \bibfield  {author} {\bibinfo {author} {\bibfnamefont {L.~M.}\ \bibnamefont
  {Janssen}}, \bibinfo {author} {\bibfnamefont {A.}~\bibnamefont {Kaiser}}, \
  and\ \bibinfo {author} {\bibfnamefont {H.}~\bibnamefont {L{\"o}wen}},\
  }\href@noop {} {\bibfield  {journal} {\bibinfo  {journal} {Sci. Rep.}\
  }\textbf {\bibinfo {volume} {7}} (\bibinfo {year} {2017})}\BibitemShut
  {NoStop}%
\bibitem [{\citenamefont {Shankar}\ \emph {et~al.}(2017)\citenamefont
  {Shankar}, \citenamefont {Bowick},\ and\ \citenamefont
  {Marchetti}}]{shankar2017topological}%
  \BibitemOpen
  \bibfield  {author} {\bibinfo {author} {\bibfnamefont {S.}~\bibnamefont
  {Shankar}}, \bibinfo {author} {\bibfnamefont {M.~J.}\ \bibnamefont {Bowick}},
  \ and\ \bibinfo {author} {\bibfnamefont {M.~C.}\ \bibnamefont {Marchetti}},\
  }\href@noop {} {\bibfield  {journal} {\bibinfo  {journal} {arXiv:1704.05424}\
  } (\bibinfo {year} {2017})}\BibitemShut {NoStop}%
\bibitem [{\citenamefont {Salbreux}\ and\ \citenamefont
  {J{\"u}licher}(2017)}]{salbreux2017mechanics}%
  \BibitemOpen
  \bibfield  {author} {\bibinfo {author} {\bibfnamefont {G.}~\bibnamefont
  {Salbreux}}\ and\ \bibinfo {author} {\bibfnamefont {F.}~\bibnamefont
  {J{\"u}licher}},\ }\href@noop {} {\bibfield  {journal} {\bibinfo  {journal}
  {Phys. Rev. E}\ }\textbf {\bibinfo {volume} {96}},\ \bibinfo {pages} {032404}
  (\bibinfo {year} {2017})}\BibitemShut {NoStop}%
\bibitem [{\citenamefont {Henkes}\ \emph {et~al.}(2017)\citenamefont {Henkes},
  \citenamefont {Marchetti},\ and\ \citenamefont
  {Sknepnek}}]{henkes2017dynamical}%
  \BibitemOpen
  \bibfield  {author} {\bibinfo {author} {\bibfnamefont {S.}~\bibnamefont
  {Henkes}}, \bibinfo {author} {\bibfnamefont {M.~C.}\ \bibnamefont
  {Marchetti}}, \ and\ \bibinfo {author} {\bibfnamefont {R.}~\bibnamefont
  {Sknepnek}},\ }\href@noop {} {\bibfield  {journal} {\bibinfo  {journal}
  {arXiv:1705.05166}\ } (\bibinfo {year} {2017})}\BibitemShut {NoStop}%
\bibitem [{\citenamefont {Duan}\ and\ \citenamefont
  {Yao}(2017)}]{duan2017curvature}%
  \BibitemOpen
  \bibfield  {author} {\bibinfo {author} {\bibfnamefont {X.}~\bibnamefont
  {Duan}}\ and\ \bibinfo {author} {\bibfnamefont {Z.}~\bibnamefont {Yao}},\
  }\href@noop {} {\bibfield  {journal} {\bibinfo  {journal} {Phys. Rev. E}\
  }\textbf {\bibinfo {volume} {95}},\ \bibinfo {pages} {062706} (\bibinfo
  {year} {2017})}\BibitemShut {NoStop}%
\bibitem [{\citenamefont {Fily}\ \emph {et~al.}(2014)\citenamefont {Fily},
  \citenamefont {Baskaran},\ and\ \citenamefont {Hagan}}]{fily2014dynamics}%
  \BibitemOpen
  \bibfield  {author} {\bibinfo {author} {\bibfnamefont {Y.}~\bibnamefont
  {Fily}}, \bibinfo {author} {\bibfnamefont {A.}~\bibnamefont {Baskaran}}, \
  and\ \bibinfo {author} {\bibfnamefont {M.~F.}\ \bibnamefont {Hagan}},\
  }\href@noop {} {\bibfield  {journal} {\bibinfo  {journal} {Soft Matter}\
  }\textbf {\bibinfo {volume} {10}},\ \bibinfo {pages} {5609} (\bibinfo {year}
  {2014})}\BibitemShut {NoStop}%
\bibitem [{\citenamefont {Fily}\ \emph {et~al.}(2015)\citenamefont {Fily},
  \citenamefont {Baskaran},\ and\ \citenamefont {Hagan}}]{fily2015dynamics}%
  \BibitemOpen
  \bibfield  {author} {\bibinfo {author} {\bibfnamefont {Y.}~\bibnamefont
  {Fily}}, \bibinfo {author} {\bibfnamefont {A.}~\bibnamefont {Baskaran}}, \
  and\ \bibinfo {author} {\bibfnamefont {M.~F.}\ \bibnamefont {Hagan}},\
  }\href@noop {} {\bibfield  {journal} {\bibinfo  {journal} {Phys. Rev. E}\
  }\textbf {\bibinfo {volume} {91}},\ \bibinfo {pages} {012125} (\bibinfo
  {year} {2015})}\BibitemShut {NoStop}%
\bibitem [{\citenamefont {Malgaretti}\ and\ \citenamefont
  {Stark}(2017)}]{malgaretti2017model}%
  \BibitemOpen
  \bibfield  {author} {\bibinfo {author} {\bibfnamefont {P.}~\bibnamefont
  {Malgaretti}}\ and\ \bibinfo {author} {\bibfnamefont {H.}~\bibnamefont
  {Stark}},\ }\href@noop {} {\bibfield  {journal} {\bibinfo  {journal} {J.
  Chem. Phys.}\ }\textbf {\bibinfo {volume} {146}},\ \bibinfo {pages} {174901}
  (\bibinfo {year} {2017})}\BibitemShut {NoStop}%
\bibitem [{\citenamefont {Beresnev}\ and\ \citenamefont
  {Nikolaevskiy}(1993)}]{1993BeNi_PhysD}%
  \BibitemOpen
  \bibfield  {author} {\bibinfo {author} {\bibfnamefont {I.~A.}\ \bibnamefont
  {Beresnev}}\ and\ \bibinfo {author} {\bibfnamefont {V.~N.}\ \bibnamefont
  {Nikolaevskiy}},\ }\href@noop {} {\bibfield  {journal} {\bibinfo  {journal}
  {Physica D}\ }\textbf {\bibinfo {volume} {66}},\ \bibinfo {pages} {1}
  (\bibinfo {year} {1993})}\BibitemShut {NoStop}%
\bibitem [{\citenamefont {Tribelsky}\ and\ \citenamefont
  {Tsuboi}(1996)}]{1996Tribelsky_PRL}%
  \BibitemOpen
  \bibfield  {author} {\bibinfo {author} {\bibfnamefont {M.~I.}\ \bibnamefont
  {Tribelsky}}\ and\ \bibinfo {author} {\bibfnamefont {K.}~\bibnamefont
  {Tsuboi}},\ }\href@noop {} {\bibfield  {journal} {\bibinfo  {journal} {Phys.
  Rev. Lett.}\ }\textbf {\bibinfo {volume} {76}},\ \bibinfo {pages} {1631}
  (\bibinfo {year} {1996})}\BibitemShut {NoStop}%
\bibitem [{\citenamefont {S{\l}omka}\ and\ \citenamefont
  {Dunkel}(2017{\natexlab{a}})}]{slomka2017geometry}%
  \BibitemOpen
  \bibfield  {author} {\bibinfo {author} {\bibfnamefont {J.}~\bibnamefont
  {S{\l}omka}}\ and\ \bibinfo {author} {\bibfnamefont {J.}~\bibnamefont
  {Dunkel}},\ }\href@noop {} {\bibfield  {journal} {\bibinfo  {journal} {Phys.
  Rev. Fluids}\ }\textbf {\bibinfo {volume} {2}},\ \bibinfo {pages} {043102}
  (\bibinfo {year} {2017}{\natexlab{a}})}\BibitemShut {NoStop}%
\bibitem [{\citenamefont {S{\l}omka}\ and\ \citenamefont
  {Dunkel}(2017{\natexlab{b}})}]{slomka2017spontaneous}%
  \BibitemOpen
  \bibfield  {author} {\bibinfo {author} {\bibfnamefont {J.}~\bibnamefont
  {S{\l}omka}}\ and\ \bibinfo {author} {\bibfnamefont {J.}~\bibnamefont
  {Dunkel}},\ }\href@noop {} {\bibfield  {journal} {\bibinfo  {journal} {Proc.
  Nat. Acad. Sci. U.S.A.}\ }\textbf {\bibinfo {volume} {114}},\ \bibinfo
  {pages} {2119} (\bibinfo {year} {2017}{\natexlab{b}})}\BibitemShut {NoStop}%
\bibitem [{\citenamefont {Burns}\ \emph {et~al.}(2017)\citenamefont {Burns},
  \citenamefont {Vasil}, \citenamefont {Oishi}, \citenamefont {Lecoanet},
  \citenamefont {Brown},\ and\ \citenamefont {Quataert}}]{dedalus2017}%
  \BibitemOpen
  \bibfield  {author} {\bibinfo {author} {\bibfnamefont {K.~J.}\ \bibnamefont
  {Burns}}, \bibinfo {author} {\bibfnamefont {G.~M.}\ \bibnamefont {Vasil}},
  \bibinfo {author} {\bibfnamefont {J.~S.}\ \bibnamefont {Oishi}}, \bibinfo
  {author} {\bibfnamefont {D.}~\bibnamefont {Lecoanet}}, \bibinfo {author}
  {\bibfnamefont {B.~P.}\ \bibnamefont {Brown}}, \ and\ \bibinfo {author}
  {\bibfnamefont {E.}~\bibnamefont {Quataert}},\ }\href@noop {} {\bibfield
  {journal} {\bibinfo  {journal} {in preparation}\ } (\bibinfo {year}
  {2017})}\BibitemShut {NoStop}%
\bibitem [{\citenamefont {Guasto}\ \emph {et~al.}(2010)\citenamefont {Guasto},
  \citenamefont {Johnson},\ and\ \citenamefont {Gollub}}]{2010Guasto}%
  \BibitemOpen
  \bibfield  {author} {\bibinfo {author} {\bibfnamefont {J.~S.}\ \bibnamefont
  {Guasto}}, \bibinfo {author} {\bibfnamefont {K.~A.}\ \bibnamefont {Johnson}},
  \ and\ \bibinfo {author} {\bibfnamefont {J.~P.}\ \bibnamefont {Gollub}},\
  }\href@noop {} {\bibfield  {journal} {\bibinfo  {journal} {Phys. Rev. Lett.}\
  }\textbf {\bibinfo {volume} {105}},\ \bibinfo {pages} {168102} (\bibinfo
  {year} {2010})}\BibitemShut {NoStop}%
\bibitem [{\citenamefont {Goldstein}\ \emph {et~al.}(2014)\citenamefont
  {Goldstein}, \citenamefont {McTavish}, \citenamefont {Moffatt},\ and\
  \citenamefont {Pesci}}]{Goldstein10062014}%
  \BibitemOpen
  \bibfield  {author} {\bibinfo {author} {\bibfnamefont {R.~E.}\ \bibnamefont
  {Goldstein}}, \bibinfo {author} {\bibfnamefont {J.}~\bibnamefont {McTavish}},
  \bibinfo {author} {\bibfnamefont {H.~K.}\ \bibnamefont {Moffatt}}, \ and\
  \bibinfo {author} {\bibfnamefont {A.~I.}\ \bibnamefont {Pesci}},\ }\href@noop
  {} {\bibfield  {journal} {\bibinfo  {journal} {Proc. Nat. Acad. Sci. U.S.A.}\
  }\textbf {\bibinfo {volume} {111}},\ \bibinfo {pages} {8339} (\bibinfo {year}
  {2014})}\BibitemShut {NoStop}%
\bibitem [{\citenamefont {Lauga}\ and\ \citenamefont
  {Powers}(2009)}]{2009LaugaPowers}%
  \BibitemOpen
  \bibfield  {author} {\bibinfo {author} {\bibfnamefont {E.}~\bibnamefont
  {Lauga}}\ and\ \bibinfo {author} {\bibfnamefont {T.~R.}\ \bibnamefont
  {Powers}},\ }\href@noop {} {\bibfield  {journal} {\bibinfo  {journal} {Rep.
  Prog. Phys.}\ }\textbf {\bibinfo {volume} {72}},\ \bibinfo {pages} {096601}
  (\bibinfo {year} {2009})}\BibitemShut {NoStop}%
\bibitem [{\citenamefont {Drescher}\ \emph {et~al.}(2011)\citenamefont
  {Drescher}, \citenamefont {Dunkel}, \citenamefont {Cisneros}, \citenamefont
  {Ganguly},\ and\ \citenamefont {Goldstein}}]{2011DrescherEtAl}%
  \BibitemOpen
  \bibfield  {author} {\bibinfo {author} {\bibfnamefont {K.}~\bibnamefont
  {Drescher}}, \bibinfo {author} {\bibfnamefont {J.}~\bibnamefont {Dunkel}},
  \bibinfo {author} {\bibfnamefont {L.~H.}\ \bibnamefont {Cisneros}}, \bibinfo
  {author} {\bibfnamefont {S.}~\bibnamefont {Ganguly}}, \ and\ \bibinfo
  {author} {\bibfnamefont {R.~E.}\ \bibnamefont {Goldstein}},\ }\href@noop {}
  {\bibfield  {journal} {\bibinfo  {journal} {Proc. Natl. Acad. Sci. U.S.A.}\
  }\textbf {\bibinfo {volume} {108}},\ \bibinfo {pages} {10940} (\bibinfo
  {year} {2011})}\BibitemShut {NoStop}%
\bibitem [{\citenamefont {Sanchez}\ \emph {et~al.}(2012)\citenamefont
  {Sanchez}, \citenamefont {Chen}, \citenamefont {DeCamp}, \citenamefont
  {Heymann},\ and\ \citenamefont {Dogic}}]{2012Sanchez_Nature}%
  \BibitemOpen
  \bibfield  {author} {\bibinfo {author} {\bibfnamefont {T.}~\bibnamefont
  {Sanchez}}, \bibinfo {author} {\bibfnamefont {D.~T.~N.}\ \bibnamefont
  {Chen}}, \bibinfo {author} {\bibfnamefont {S.~J.}\ \bibnamefont {DeCamp}},
  \bibinfo {author} {\bibfnamefont {M.}~\bibnamefont {Heymann}}, \ and\
  \bibinfo {author} {\bibfnamefont {Z.}~\bibnamefont {Dogic}},\ }\href@noop {}
  {\bibfield  {journal} {\bibinfo  {journal} {Nature}\ }\textbf {\bibinfo
  {volume} {491}},\ \bibinfo {pages} {431} (\bibinfo {year}
  {2012})}\BibitemShut {NoStop}%
\bibitem [{\citenamefont {Scriven}(1960{\natexlab{a}})}]{scriven1960dynamics}%
  \BibitemOpen
  \bibfield  {author} {\bibinfo {author} {\bibfnamefont {L.~E.}\ \bibnamefont
  {Scriven}},\ }\href@noop {} {\bibfield  {journal} {\bibinfo  {journal} {Chem.
  Eng. Sci.}\ }\textbf {\bibinfo {volume} {12}},\ \bibinfo {pages} {98}
  (\bibinfo {year} {1960}{\natexlab{a}})}\BibitemShut {NoStop}%
\bibitem [{\citenamefont {Aris}(1989)}]{aris1989vectors}%
  \BibitemOpen
  \bibfield  {author} {\bibinfo {author} {\bibfnamefont {R.}~\bibnamefont
  {Aris}},\ }\href@noop {} {\emph {\bibinfo {title} {Vectors, tensors and the
  basic equations of fluid mechanics}}}\ (\bibinfo  {publisher} {Dover
  Publications, Inc.},\ \bibinfo {address} {New York},\ \bibinfo {year}
  {1989})\BibitemShut {NoStop}%
\bibitem [{Note1()}]{Note1}%
  \BibitemOpen
  \bibinfo {note} {G. M. Vasil and M. G. P. Cassell, in
  preparation.}\BibitemShut {Stop}%
\bibitem [{SM()}]{SM}%
  \BibitemOpen
  \href@noop {} {}\bibinfo {note} {See Supporting Material, which contains
  Refs.~\cite{schwarz1995hodge,scriven1960,delay2007tt,besse2007einstein,arnold1999topological}.}\BibitemShut
  {Stop}%
\bibitem [{\citenamefont {Prandl}\ \emph {et~al.}(1996)\citenamefont {Prandl},
  \citenamefont {Schiebel},\ and\ \citenamefont {Wulf}}]{prandl1996recursive}%
  \BibitemOpen
  \bibfield  {author} {\bibinfo {author} {\bibfnamefont {W.}~\bibnamefont
  {Prandl}}, \bibinfo {author} {\bibfnamefont {P.}~\bibnamefont {Schiebel}}, \
  and\ \bibinfo {author} {\bibfnamefont {K.}~\bibnamefont {Wulf}},\ }\href@noop
  {} {\bibfield  {journal} {\bibinfo  {journal} {Acta Crystallogr. Sect. A}\
  }\textbf {\bibinfo {volume} {52}},\ \bibinfo {pages} {171} (\bibinfo {year}
  {1996})}\BibitemShut {NoStop}%
\bibitem [{\citenamefont {Waleffe}(2001)}]{2001Waleffe_JFM}%
  \BibitemOpen
  \bibfield  {author} {\bibinfo {author} {\bibfnamefont {F.}~\bibnamefont
  {Waleffe}},\ }\href@noop {} {\bibfield  {journal} {\bibinfo  {journal} {J.
  Fluid Mech.}\ }\textbf {\bibinfo {volume} {435}},\ \bibinfo {pages} {93}
  (\bibinfo {year} {2001})}\BibitemShut {NoStop}%
\bibitem [{\citenamefont {Waleffe}(2009)}]{2009Waleffe}%
  \BibitemOpen
  \bibfield  {author} {\bibinfo {author} {\bibfnamefont {F.}~\bibnamefont
  {Waleffe}},\ }in\ \href@noop {} {\emph {\bibinfo {booktitle} {Turbulence and
  Interactions}}},\ \bibinfo {series} {Notes on Numerical Fluid Mechanics and
  Multidisciplinary Design}, Vol.\ \bibinfo {volume} {105},\ \bibinfo {editor}
  {edited by\ \bibinfo {editor} {\bibfnamefont {M.}~\bibnamefont {Deville}},
  \bibinfo {editor} {\bibfnamefont {T.~H.}\ \bibnamefont {L$\hat{\text{e}}$}},
  \ and\ \bibinfo {editor} {\bibfnamefont {P.}~\bibnamefont {Sagaut}}}\
  (\bibinfo  {publisher} {Springer},\ \bibinfo {address} {Berlin, Heidelberg},\
  \bibinfo {year} {2009})\BibitemShut {NoStop}%
\bibitem [{\citenamefont {Jiang}\ \emph {et~al.}(2011)\citenamefont {Jiang},
  \citenamefont {Machiraju},\ and\ \citenamefont
  {Thompson}}]{jiang2005detection}%
  \BibitemOpen
  \bibfield  {author} {\bibinfo {author} {\bibfnamefont {M.}~\bibnamefont
  {Jiang}}, \bibinfo {author} {\bibfnamefont {R.}~\bibnamefont {Machiraju}}, \
  and\ \bibinfo {author} {\bibfnamefont {D.}~\bibnamefont {Thompson}},\ }in\
  \href@noop {} {\emph {\bibinfo {booktitle} {The Visualization Handbook}}},\
  \bibinfo {editor} {edited by\ \bibinfo {editor} {\bibfnamefont {C.~D.}\
  \bibnamefont {Hansen}}\ and\ \bibinfo {editor} {\bibfnamefont {C.~R.}\
  \bibnamefont {Johnson}}}\ (\bibinfo  {publisher} {Academic Press},\ \bibinfo
  {year} {2011})\BibitemShut {NoStop}%
\bibitem [{\citenamefont {Boffetta}\ and\ \citenamefont
  {Ecke}(2012)}]{boffetta2012two}%
  \BibitemOpen
  \bibfield  {author} {\bibinfo {author} {\bibfnamefont {G.}~\bibnamefont
  {Boffetta}}\ and\ \bibinfo {author} {\bibfnamefont {R.~E.}\ \bibnamefont
  {Ecke}},\ }\href@noop {} {\bibfield  {journal} {\bibinfo  {journal} {Annu.
  Rev. Fluid Mech.}\ }\textbf {\bibinfo {volume} {44}},\ \bibinfo {pages} {427}
  (\bibinfo {year} {2012})}\BibitemShut {NoStop}%
\bibitem [{\citenamefont {Kraichnan}(1967)}]{kraichnan1967inertial}%
  \BibitemOpen
  \bibfield  {author} {\bibinfo {author} {\bibfnamefont {R.~H.}\ \bibnamefont
  {Kraichnan}},\ }\href@noop {} {\bibfield  {journal} {\bibinfo  {journal}
  {Phys. Fluids}\ }\textbf {\bibinfo {volume} {10}},\ \bibinfo {pages} {1417}
  (\bibinfo {year} {1967})}\BibitemShut {NoStop}%
\bibitem [{\citenamefont {Tang}\ and\ \citenamefont
  {Orszag}(1978)}]{tang1978two}%
  \BibitemOpen
  \bibfield  {author} {\bibinfo {author} {\bibfnamefont {C.-M.}\ \bibnamefont
  {Tang}}\ and\ \bibinfo {author} {\bibfnamefont {S.~A.}\ \bibnamefont
  {Orszag}},\ }\href@noop {} {\bibfield  {journal} {\bibinfo  {journal} {J.
  Fluid Mech.}\ }\textbf {\bibinfo {volume} {87}},\ \bibinfo {pages} {305}
  (\bibinfo {year} {1978})}\BibitemShut {NoStop}%
\bibitem [{\citenamefont {Kraichnan}\ and\ \citenamefont
  {Montgomery}(1980)}]{kraichnan1980two}%
  \BibitemOpen
  \bibfield  {author} {\bibinfo {author} {\bibfnamefont {R.~H.}\ \bibnamefont
  {Kraichnan}}\ and\ \bibinfo {author} {\bibfnamefont {D.}~\bibnamefont
  {Montgomery}},\ }\href@noop {} {\bibfield  {journal} {\bibinfo  {journal}
  {Rep. Prog. Phys.}\ }\textbf {\bibinfo {volume} {43}},\ \bibinfo {pages}
  {547} (\bibinfo {year} {1980})}\BibitemShut {NoStop}%
\bibitem [{\citenamefont {Sukoriansky}\ \emph {et~al.}(2002)\citenamefont
  {Sukoriansky}, \citenamefont {Galperin},\ and\ \citenamefont
  {Dikovskaya}}]{sukoriansky2002universal}%
  \BibitemOpen
  \bibfield  {author} {\bibinfo {author} {\bibfnamefont {S.}~\bibnamefont
  {Sukoriansky}}, \bibinfo {author} {\bibfnamefont {B.}~\bibnamefont
  {Galperin}}, \ and\ \bibinfo {author} {\bibfnamefont {N.}~\bibnamefont
  {Dikovskaya}},\ }\href@noop {} {\bibfield  {journal} {\bibinfo  {journal}
  {Phys. Rev. Lett.}\ }\textbf {\bibinfo {volume} {89}},\ \bibinfo {pages}
  {124501} (\bibinfo {year} {2002})}\BibitemShut {NoStop}%
\bibitem [{\citenamefont {Biferale}\ \emph {et~al.}(2012)\citenamefont
  {Biferale}, \citenamefont {Musacchio},\ and\ \citenamefont
  {Toschi}}]{biferale2012inverse}%
  \BibitemOpen
  \bibfield  {author} {\bibinfo {author} {\bibfnamefont {L.}~\bibnamefont
  {Biferale}}, \bibinfo {author} {\bibfnamefont {S.}~\bibnamefont {Musacchio}},
  \ and\ \bibinfo {author} {\bibfnamefont {F.}~\bibnamefont {Toschi}},\
  }\href@noop {} {\bibfield  {journal} {\bibinfo  {journal} {Phys. Rev. Lett.}\
  }\textbf {\bibinfo {volume} {108}},\ \bibinfo {pages} {164501} (\bibinfo
  {year} {2012})}\BibitemShut {NoStop}%
\bibitem [{\citenamefont {Schwarz}(1995)}]{schwarz1995hodge}%
  \BibitemOpen
  \bibfield  {author} {\bibinfo {author} {\bibfnamefont {G.}~\bibnamefont
  {Schwarz}},\ }\href@noop {} {\emph {\bibinfo {title} {Hodge Decomposition - A
  method for solving boundary value problems}}}\ (\bibinfo  {publisher}
  {Springer-Verlag},\ \bibinfo {year} {1995})\BibitemShut {NoStop}%
\bibitem [{\citenamefont {Scriven}(1960{\natexlab{b}})}]{scriven1960}%
  \BibitemOpen
  \bibfield  {author} {\bibinfo {author} {\bibfnamefont {L.}~\bibnamefont
  {Scriven}},\ }\href@noop {} {\bibfield  {journal} {\bibinfo  {journal} {Chem.
  Eng. Sci.}\ }\textbf {\bibinfo {volume} {12}},\ \bibinfo {pages} {98}
  (\bibinfo {year} {1960}{\natexlab{b}})}\BibitemShut {NoStop}%
\bibitem [{\citenamefont {Delay}(2007)}]{delay2007tt}%
  \BibitemOpen
  \bibfield  {author} {\bibinfo {author} {\bibfnamefont {E.}~\bibnamefont
  {Delay}},\ }\href@noop {} {\bibfield  {journal} {\bibinfo  {journal}
  {Manuscripta Math.}\ }\textbf {\bibinfo {volume} {123}},\ \bibinfo {pages}
  {147} (\bibinfo {year} {2007})}\BibitemShut {NoStop}%
\bibitem [{\citenamefont {Besse}(2007)}]{besse2007einstein}%
  \BibitemOpen
  \bibfield  {author} {\bibinfo {author} {\bibfnamefont {A.~L.}\ \bibnamefont
  {Besse}},\ }\href@noop {} {\emph {\bibinfo {title} {Einstein manifolds}}}\
  (\bibinfo  {publisher} {Springer Science \& Business Media},\ \bibinfo {year}
  {2007})\BibitemShut {NoStop}%
\bibitem [{\citenamefont {Arnold}\ and\ \citenamefont
  {Khesin}(1999)}]{arnold1999topological}%
  \BibitemOpen
  \bibfield  {author} {\bibinfo {author} {\bibfnamefont {V.~I.}\ \bibnamefont
  {Arnold}}\ and\ \bibinfo {author} {\bibfnamefont {B.~A.}\ \bibnamefont
  {Khesin}},\ }\href@noop {} {\emph {\bibinfo {title} {Topological methods in
  hydrodynamics}}},\ Vol.\ \bibinfo {volume} {125}\ (\bibinfo  {publisher}
  {Springer Science \& Business Media},\ \bibinfo {year} {1999})\BibitemShut
  {NoStop}%
\end{thebibliography}
\end{document}